\documentclass[iop,apj,tighten]{emulateapj}

\usepackage{apjfonts} 
\usepackage{amsmath,amstext}
\usepackage{verbatim} 
\usepackage[breaklinks,colorlinks,citecolor=blue,urlcolor=magenta,linkcolor=blue]{hyperref}
\usepackage[toc,page]{appendix}
\usepackage{ulem} 
\usepackage{bm} 

\defcitealias{Gordon:2003}{G03}
\defcitealias{WD:2001}{WD01}
\defcitealias{Fitzpatrick:1999}{F99}
\defcitealias{DeMarchi:2014fe}{DM14a}
\defcitealias{DeMarchi:2014en}{DM14b}
\defcitealias{DeMarchi:2016}{DM16}

\newcommand{\angstrom }{\mbox{\normalfont\AA}} 
\newcommand{\about}{\textasciitilde} 
\newcommand{\fsf}{\textit{F475W}} 
\newcommand{\eof}{\textit{F814W}} 

\shorttitle{SMIDGE SMC Extinction with Red Clump Stars}
\shortauthors{Yanchulova Merica-Jones et al.}

\begin{document}

\title{The Small Magellanic Cloud Investigation of Dust and Gas Evolution (SMIDGE): The Dust Extinction Curve from Red Clump Stars}
\author{Petia Yanchulova Merica-Jones\altaffilmark{1}, Karin M. Sandstrom\altaffilmark{1}, L. Clifton Johnson\altaffilmark{1}, Julianne Dalcanton\altaffilmark{2}, Andrew E. Dolphin\altaffilmark{3}, Karl Gordon\altaffilmark{4, 5}, Julia Roman-Duval\altaffilmark{4}, Daniel R. Weisz\altaffilmark{6}, Benjamin F. Williams\altaffilmark{2}}
\affil{$^1$Center for Astrophysics and Space Sciences, Department of Physics, University of California, 9500 Gilman Drive, La Jolla, San Diego, CA 92093, USA}
\affil{ $^2$Department of Astronomy, University of Washington, Box 351580, Seattle, WA 98195, USA}
\affil{$^3$Raytheon; 1151 E. Hermans Road, Tucson, AZ 85756, USA}
\affil{$^4$Space Telescope Science Institute, 3700 San Martin Drive, Baltimore, MD 21218, USA}
\affil{$^5$Sterrenkundig Observatorium, Universiteit Gent, Gent, Belgium}
\affil{$^6$Department of Astronomy, University of California, 501 Campbell Hall \#3411, Berkeley, CA 94720-3411, USA}

\begin{abstract}
We use \textit{Hubble Space Telescope (HST)} observations of red clump stars taken as part of the Small Magellanic Cloud Investigation of Dust and Gas Evolution (SMIDGE) program to measure the average dust extinction curve in a \about 200 pc $\times$ 100 pc region in the southwest bar of the Small Magellanic Cloud (SMC).  The rich information provided by our 8-band ultra-violet through near-infrared photometry allows us to model the color-magnitude diagram of the red clump accounting for the extinction curve shape, a log-normal distribution of $A_{V}$, and the depth of the stellar distribution along the line of sight.  We measure an extinction curve with $R_{475} = A_{475}/(A_{475}-A_{814})$ = 2.65 $\pm$ 0.11.  This measurement is significantly larger than the equivalent values of published Milky Way $R_{V}$ = 3.1 ($R_{475} = 1.83$) and SMC Bar $R_{V}$ = 2.74 ($R_{475} = 1.86$) extinction curves.  Similar extinction curve offsets in the Large Magellanic Cloud (LMC) have been interpreted as the effect of large dust grains.  We demonstrate that the line-of-sight depth of the SMC (and LMC) introduces an apparent ``gray'' contribution to the extinction curve inferred from the morphology of the red clump.  We show that no gray dust component is needed to explain extinction curve measurements when a full-width half-max depth of 10 $\pm$ 2 kpc in the stellar distribution of the SMC (5 $\pm$ 1 kpc for the LMC) is considered, which agrees with recent studies of Magellanic Cloud stellar structure. The results of our work demonstrate the power of broad-band HST imaging for simultaneously constraining dust and galactic structure outside the Milky Way.
\end{abstract}


\section{Introduction} \label{sec:intro}

\indent Studying dust and its extinction is essential for our interpretation of the observational properties and evolution of galaxies. By examining dust extinction in the Small Magellanic Cloud, we can obtain a high-resolution picture of a low metallicity environment  of $1/5-1/8 \;Z_{\odot}$ \citep{Dufour:1984,RussellDopita:1992,Kurt:1999,Lee:2005,Rolleston:1999,Rolleston:2003} at a distance of 62 kpc \citep{Scowcroft:2016hm}. SMC-like extinction is widely used to correct for the effects of dust in low metallicity or high redshift galaxies \citep{Noll:2005, Galliano:2005, Cignoni:2009, Glatt:2008, Sabbi:2009}.  Currently, however, there are only a handful of measurements of the extinction curve in the SMC.

Extinction curve measurements exist for only about ten individual SMC sight lines towards O and B stars derived from UV spectroscopy \citep{Lequeux:1982, Prevot:1984, Gordon:1998, Gordon:2003, MaizApellaniz:2012}. These few sight lines allow only low-number statistics of extinction and dust properties in the Small Magellanic Cloud while they also tend to probe the vicinity of star-forming regions which may alter the dust grains in their surroundings due to the intense UV radiation fields. A more thorough appreciation of the extinction curve shape and its possible variations can be gained through large-scale multiwavelength studies which can reveal features difficult to reliably detect by observing a handful of individual stars \citep[e.g.][]{Schlafly:2016}. To fully understand dust extinction in a low-metallicity environment it is therefore critical to study the SMC on a wider scale to obtain a more representative sample of extinction curves.

\begin{figure*} 
  \centering 
    \includegraphics[width=\textwidth]{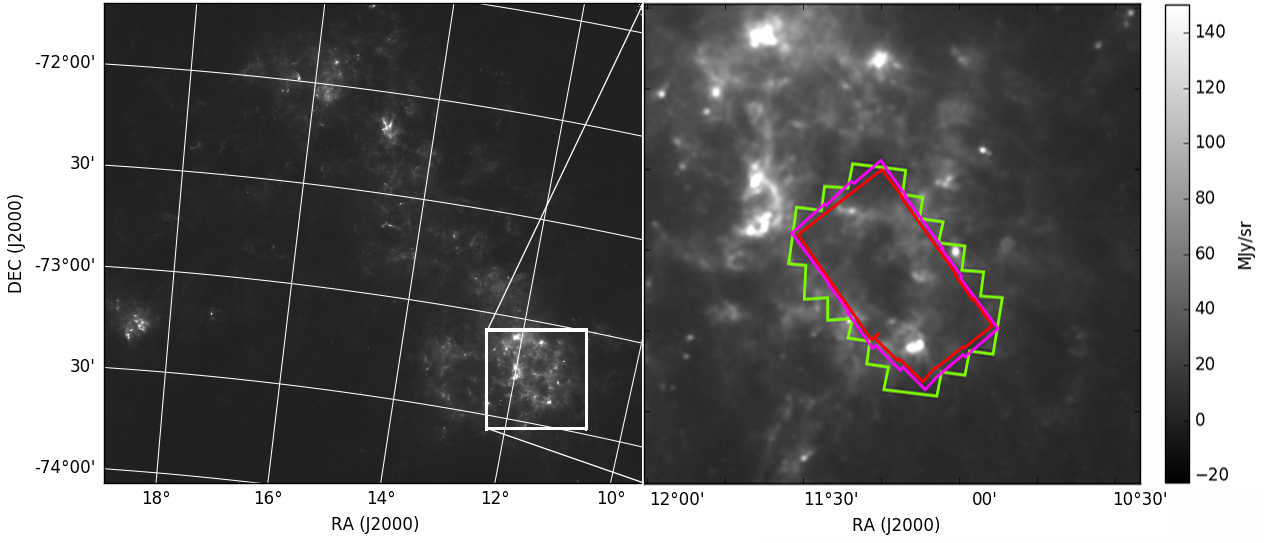}
  \caption{Location of the SMIDGE region in the southwest bar of the Small Magellanic Cloud superimposed on a Herschel 250 $\mu$m map. The imaging footprint of \textit{Hubble Space Telescope's} cameras are shown in green (ACS/WFC), magenta (WFC3/UVIS), and red (WFC3/IR).}
  \label{fig:SMCregions}
\end{figure*}

It is immediately clear from existing SMC extinction curve measurements that SMC-like dust is distinct from Milky Way dust.  Most extinction curves in the Milky Way (MW) are well-described by an empirical relationship based on one parameter, $R_{V} = A_{V}/E(B-V) = A_{V}/(A_{B} - A_{V})$, representing the ratio of total to selective extinction at optical wavelengths \citep{CCM:1989} and serving as a proxy for the average dust grain size along a sight line. Curves outside the Galaxy show deviations from this relationship, including all measured curves to date in the SMC \citep{Gordon:2003,Cartledge:2005,MaizApellaniz:2012}. Particular differences from MW-type dust extinction including a steeper far-UV rise and/or a weaker 2175 \angstrom\ bump are evident near the 30 Doradus region in the Large Magellanic Cloud (LMC) \citep{ClaytonMartin:1985,Fitzpatrick:1985} and in the star-forming bar of the SMC \citep{Prevot:1984}.

The pair method \citep{Trumpler:1930, Massa:1983, Cardelli:1992} has produced the majority of extinction curves in the SMC. This method measures extinction as a function of wavelength by comparing the spectra of a reddened and an unreddened star of approximately the same spectral type. In addition to being limited to only a few lines of sight, the pair method requires high signal-to-noise ultraviolet spectroscopy which additionally places a strict limit on the brightness of stars which can be studied.  Modifications to the pair method can be made to increase the efficiency of the technique. These include using photometric measurements instead of spectroscopy and using theoretical predictions for the star's color and magnitude rather than a specific matched comparison star. In this study we use the latter approach and produce color-magnitude diagrams (CMDs) for thousands of stars allowing us to simultaneously measure the stars' displacement due to dust from a known unreddened location. Measuring this effect, which is evident across photometric bands ranging in wavelength from ultraviolet to infrared, allows us to determine the extinction curve shape.

To measure a star's displacement in CMD space due to dust extinction, we must have an idea of its initial location.  The red giant clump, or red clump (RC), population is an ideal target for such techniques due to the narrow intrinsic distribution of the stars in CMD space \citep{GirardiSalaris:2001,Girardi:2016}. RC stars are mostly low-mass 1-12 Gyr K giants in their He-burning phase found at the red end of the horizontal branch occupying a compact region on a CMD.  Due to variable reddening by dust, the RC appears as an extended sequence stretching toward fainter magnitudes and redder colors from the unreddened RC location. By measuring the vertical displacement of the reddened RC feature, it is possible to find the dust extinction in magnitudes, $A_{\lambda}$. From the slope of the reddened RC, one can also measure the value of the ratio of the absolute to selective extinction, $R_{\lambda} = A_{\lambda}/(A_{\lambda \prime} - A_{\lambda \prime\prime})$, where $A_{\lambda \prime} - A_{\lambda \prime\prime}$ is the extinction in a chosen color combination. The selective extinction is traditionally measured in $B$ and $V$ filters (at 4405 {\AA} and 5470 {\AA}) giving $A_{B} - A_{V} = E(B-V)$ and a corresponding $R_{V} = A_{V}/E(B-V)$. This technique can be applied with a variety of different color combinations to study the extinction curve shape.

\begin{figure*} 
  \centering
   \includegraphics[width=\textwidth]
   {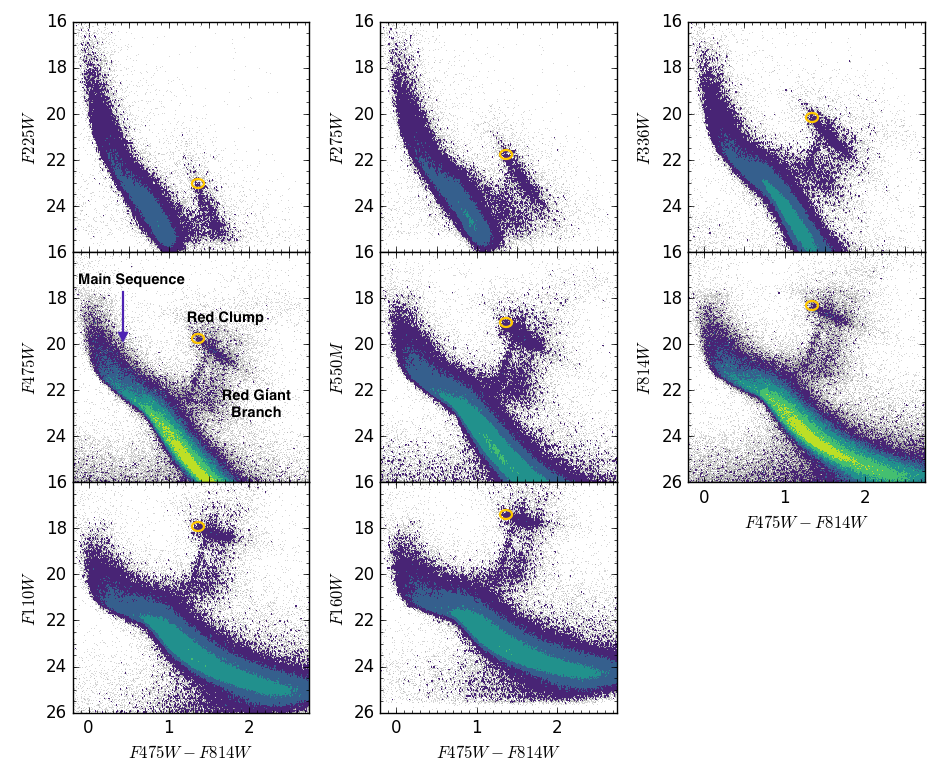}
   \caption{Color-magnitude diagrams (CMDs) for the eight SMIDGE photometric bands plotted in the optical $\fsf-\eof$ color after applying the culling values in Table \ref{tab:culltable}. Stellar density contours range from 2 to 1200 stars decimag$^{-2}$. The main features are shown in \fsf's CMD. The red clump population is seen as a streak above and somewhat parallel to the main sequence where the orange ellipse indicates its theoretical unreddened location determined as described in Sec. \ref{sec:UnredRClocation} with values listed in Table \ref{tab:unredRCloc}. The red giant branch can be distinguished by its bimodal appearance rising almost vertically between the main sequence and the red clump. Both the extended red clump streak and the doubled red giant branch are consequences of dust extinction.}
   \label{fig:cmdsAllBands}
\end{figure*}

The application of the red clump method to measure extinction has been used in the Milky Way \citep{Nataf:2013} and the LMC \citep[][hereafter DM14a, DM14b, and DM16]{DeMarchi:2014fe, DeMarchi:2014en, DeMarchi:2016}. Some of these previous studies have led to unexpected results. \citetalias{DeMarchi:2014fe} first present the red clump method (subsequently employed by \citetalias{DeMarchi:2014en} and \citetalias{DeMarchi:2016}) to study the shape of the extinction curve using multi-band photometry in LMC's Tarantula Nebula. They find $R_{V}$ values of 5.6 $\pm$ 0.3 and 4.5 $\pm$ 0.2 for R136 and 30 Doradus, respectively, indicating curves with a steeper optical reddening vector than the diffuse Galactic interstellar medium (ISM) of $R_{V}=3.1$ \citep{CCM:1989} and LMC's average of $R_{V}$=3.4 \citep[][hereafter G03]{Gordon:2003}. The authors attribute this result to the presence of ``gray'' extinction (i.e. extinction with small changes in color). ``Gray'' UV/optical extinction curves with high values of $R$ are attributed to the presence of large dust grains \citep{Strom:1971,Dunkin:1998}.  Indeed, high $R_{V}\sim 5.5$ is observed in the dense Galactic ISM, where dust grain growth is thought to have occurred \citep{CCM:1989,WD:2001}. \citetalias{DeMarchi:2014en} and \citetalias{DeMarchi:2016} argue that the observed LMC extinction curve with high $R_{V}$ is indicative of large grains injected into the ISM by recent Type II supernova explosions.

Here we present the results of a study using \textit{Hubble Space Telescope (HST)} multi-band observations of the SMC to characterize the average extinction curve at a range of wavelengths and over a large region of the galaxy.  Our survey, named the Small Magellanic Cloud Investigation of Dust and Gas Evolution (SMIDGE), focuses on a 100 pc $\times$ 200 pc region in the southwest bar of the SMC with observations spanning the ultraviolet (UV) to near-infrared (NIR) wavelength range. Similarly to the work of \citetalias{DeMarchi:2016}, we perform an analysis of HST observations of the 30 Doradus region in the LMC.  In addition to \citetalias{DeMarchi:2016}'s study, we also aim to explain how the SMC's and the LMC's depth along the line of sight impacts the extinction curve shape in the region.  We present the SMIDGE data in Section~\ref{sec:obs}.  In Section \ref{sec:cmdfitting} we give details about how the extinction curve can be measured from red clump stars and present a model of how the depth along the line of sight of the SMC impacts extinction curves measurements.  In Section ~\ref{sec:extinctionresults} we quantify the effect of the SMC and LMC depth along the line of sight on existing observed and model extinction curves, and test the agreement between the latter and the newly-derived extinction curves.  We then compare our results to other extinction curve measurements.  In Section \ref{sec:discussion} we discuss the significance of our results for the extinction curve shape in the SMC and we conclude in Section \ref{sec:conclusions}.

\section{Data}\label{sec:obs}

\begin{deluxetable}{lcc}
\tabletypesize {\small} 
\tablecolumns{3}
\tablecaption{SMIDGE quality cuts}
\tablehead{ \multicolumn{1}{l}{ Camera} &
\multicolumn{1}{c}{Sharpness} &
\multicolumn{1}{c}{Crowding} }
\startdata
 UVIS & 0.15 & 1.3\\
 ACS & 0.2 & 2.25\\
 IR & 0.15 & 2.25
\enddata
\tablecomments{For details see Section 2.3 of \cite{Williams:2014}}
\label{tab:culltable}
\end{deluxetable}

We use imaging obtained by the SMIDGE survey (GO 13659) covering an area of \about  200 pc $\times$ 100 pc in the southwest bar region of the Small Magellanic Cloud. The location of the imaging footprint is displayed in Figure \ref{fig:SMCregions}. The data were obtained by the \textit{Hubble Space Telescope's} Advanced Camera for Surveys (ACS) Wide Field Camera (WFC), and the Wide Field Camera 3 (WFC3) instrument's infrared (IR) and ultraviolet-optical (UVIS) imagers. The following broad-band filters were used, covering the wavelength range 0.24 - 1.5 $\mu$m: \textit{F225W}, \textit{F275W} and \textit{F336W} from \textit{WFC3/UVIS}; \fsf, \textit{F550M}, \textit{F658N}, and \eof from \textit{ACS/WFC}; and \textit{F110W} and \textit{F160W} from \textit{WFC3/IR}. We omit the $H\alpha$ narrow-band filter \textit{F658N} for the purposes of this study.

We perform PSF-fitting photometry using the DOLPHOT package, an updated version of HSTPHOT \citep{Dolphin:2000}.  We follow analysis procedures similar to those used by the Panchromatic Hubble Andromeda Treasury (PHAT) survey, as described by \citet{Williams:2014}.  Namely, we follow the same methodologies and processing code for image alignment, cosmic ray and artifact rejection, and DOLPHOT execution.  SMIDGE processing differs in a number of ways: we use a \texttt{TinyTim}-based PSF library, photometer full-depth stacks of all images simultaneously, and improve image distortion corrections.  Complete details about the survey's observations, reduction, data quality, and catalog presentation will be outlined in the upcoming SMIDGE survey paper (Sandstrom et al., in preparation).

We have made a number of culls on the photometry catalogs to eliminate low-quality measurements and obtain color-magnitude diagrams (CMDs) with well-defined features. Table \ref{tab:culltable} lists the sharpness and crowding culling values we have used to obtain \texttt{gst} catalogs.  We plot the CMDs for the SMIDGE survey in Figure \ref{fig:cmdsAllBands}. We note that a red leak in \textit{HST's WFC3/UV F225W} affects our photometry of the red clump. This is only an issue for the UV-faint red clump stars in \textit{F225W} where the red leak can cause a significant offset in the photometry.  Although we present the derived values of the reddening vector for \textit{F225W}, we suggest caution in interpreting these values.

\begin{figure*} 
  \centering 
    \includegraphics[width=0.48\textwidth]{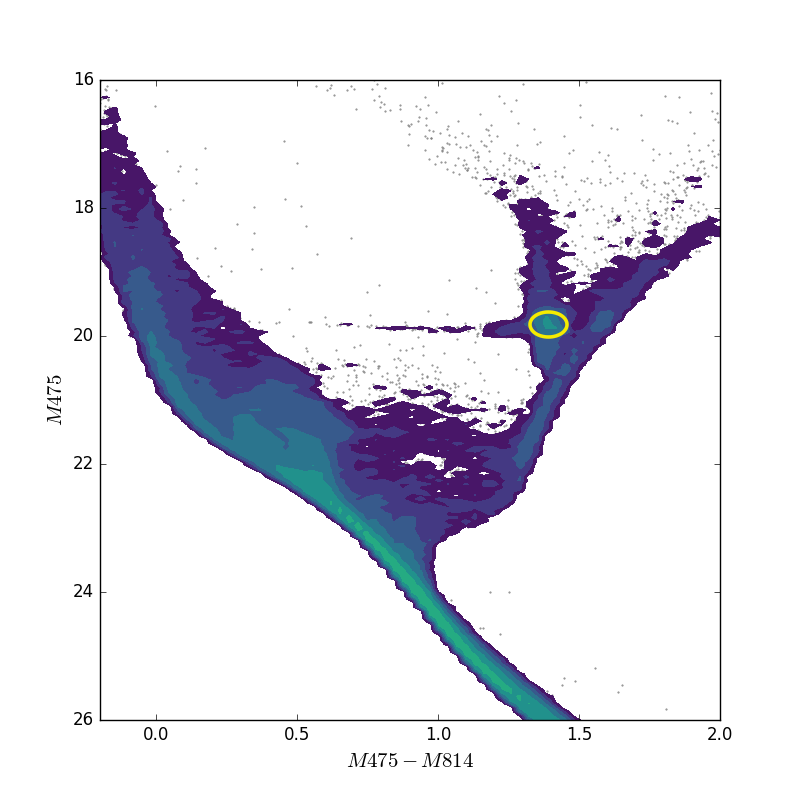}
    \hspace{0.1cm}
    \includegraphics[width=0.48\textwidth]{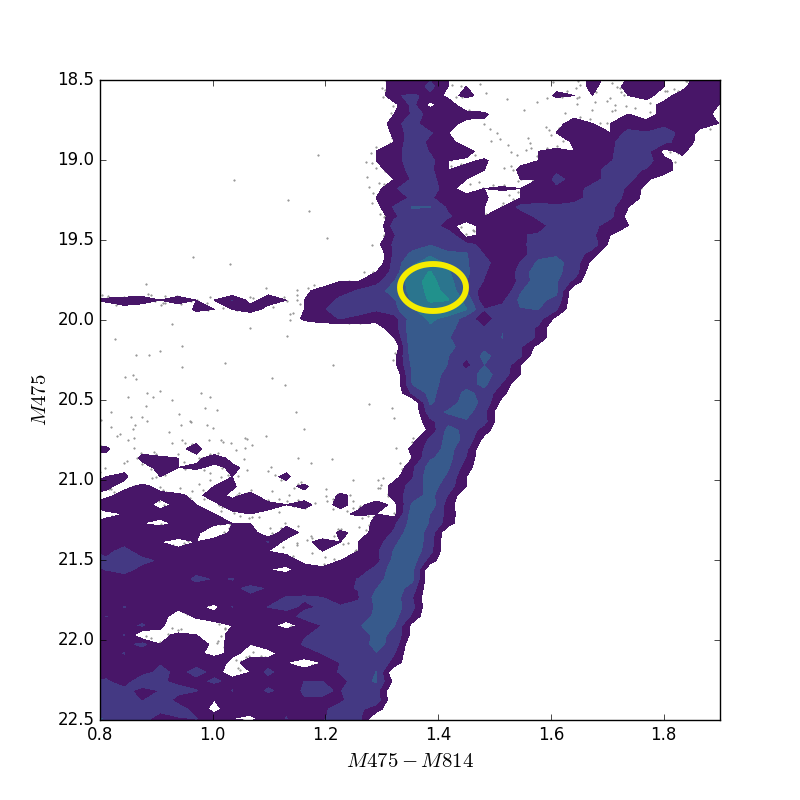}
   \caption{$\fsf$ synthetic CMD with the unreddened red clump location indicated by the yellow ellipse in both panels. The contours are logarithmic in number of stars per decimag. The selection of the unreddened red clump is based on parameters specific to the Small Magellanic Cloud as explained in Sec. \ref{sec:UnredRClocation}. For clarity, the width and height of the ellipse represent twice the size of the unreddened red clump intrinsic spread in color and magnitude.  Synthetic CMDs for the rest of the SMIDGE filters are plotted in the Appendix. }
   \label{fig:synthcmd}
\end{figure*}

\begin{deluxetable}{lcc} 
\tabletypesize{\scriptsize}
\tablecolumns{3}
\tablecaption{SMC Red Clump Properties}
\tablehead{ \multicolumn{1}{l}{ Filter} &
\multicolumn{1}{c}{Theoretical $RC_{Mag}$} &
\multicolumn{1}{c}{$\sigma_{Mag}$} }
\startdata
 \textit{F225W} &  23.22 & 0.21 \\
 \textit{F275W} &  21.83 & 0.15\\
 \textit{F336W} &  20.32 & 0.11\\
 \fsf &  19.78 & 0.09\\
 \textit{F550W} &  19.22 & 0.09\\
 \eof &  18.40 & 0.09\\
 \textit{F110W} &  17.96 & 0.09\\
 \textit{F160W} &  17.44 & 0.09
\enddata
\tablecomments{ Unreddened red clump location from synthetic CMD. The mean $\fsf-\eof$ color for all photometric bands is 1.38 $\pm$ 0.059. This includes MW foreground reddening toward the SMC of A$_{V}=0.18$ and R$_{V}=3.1$, and a distance modulus of 18.96.}
\label{tab:unredRCloc}
\end{deluxetable}

To analyze the extinction curve in the 30 Doradus region in the Large Magellanic Cloud, we use data from the Hubble Tarantula Treasury Project (HTTP) survey. The survey data, described in Section 2 of \citet{Sabbi:2016}, encompasses ultraviolet to infrared wavelengths and is obtained by the \textit{Hubble Space Telescope's} ACS and WFC3 instruments in the set of \textit{HST's F275W, F336W, F555W, F658N, F775A, F775U, F110W}, and \textit{F160W} filters, where \textit{F775A} is ACS/WFC \textit{F775W}, and \textit{F775U} is WFC3/UVIS \textit{F775W}. We choose the ACS/WFC \textit{F775W} filter due to the larger number of sources in this passband, and we omit the narrow-band filter \textit{F658N} as we do for the SMIDGE extinction curve analysis.  Further survey details about the observations, data reduction and quality, and catalog presentation are discussed in \citet{Sabbi:2016}.

\section{Extinction from Red Clump Stars}
\label{sec:cmdfitting}
\subsection{Overview \label{sec:RCextOverview}}

\begin{figure*}
  \centering
    \includegraphics[width=\textwidth]{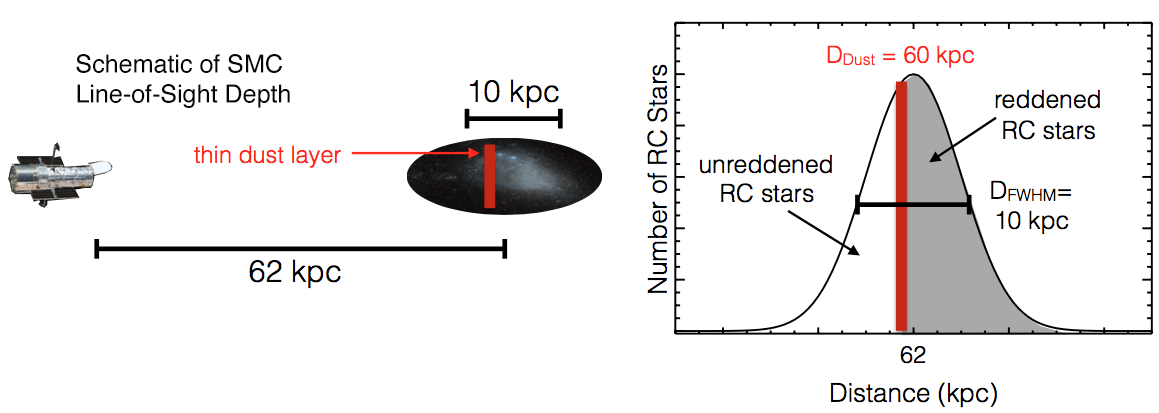}
   \caption{ Example of the stellar and dust distribution in our model with parameters corresponding to the example in Figure~\ref{fig:toymodel}. The bimodality of the red giant branch leads us to conclude that the dust is located in a thin layer relative to the stars leading to either unreddened or reddened red clump stars (see Sec.\ref{sec:ModelDust}).}
   \label{fig:distributionLOSdepth}
\end{figure*}

\begin{figure*} 
	\centering	\includegraphics[width=0.49\textwidth]
{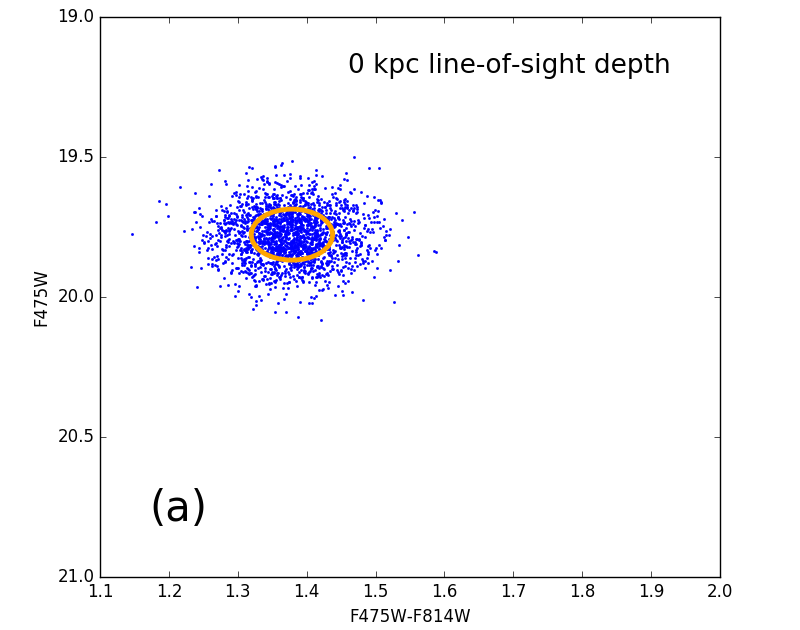}
\includegraphics[width=0.49\textwidth]
{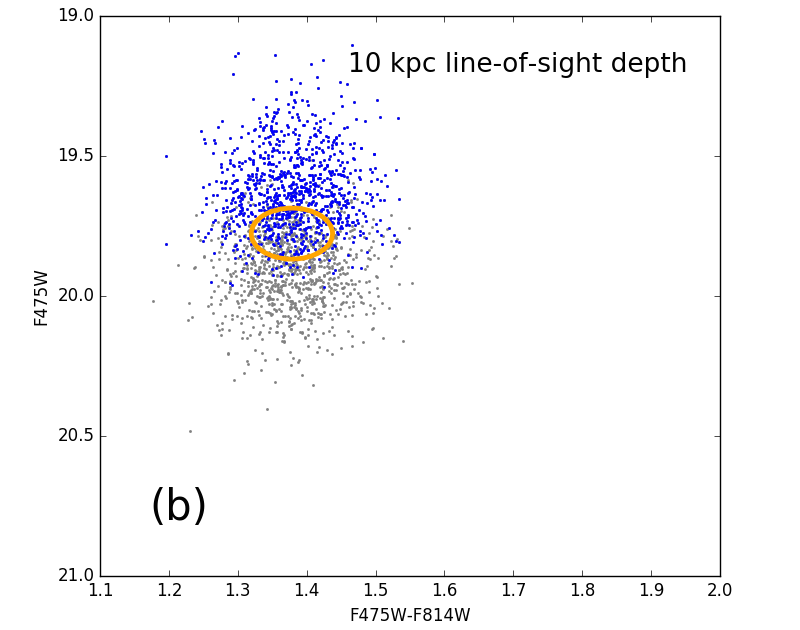}
	\centering	\includegraphics[width=0.49\textwidth]
{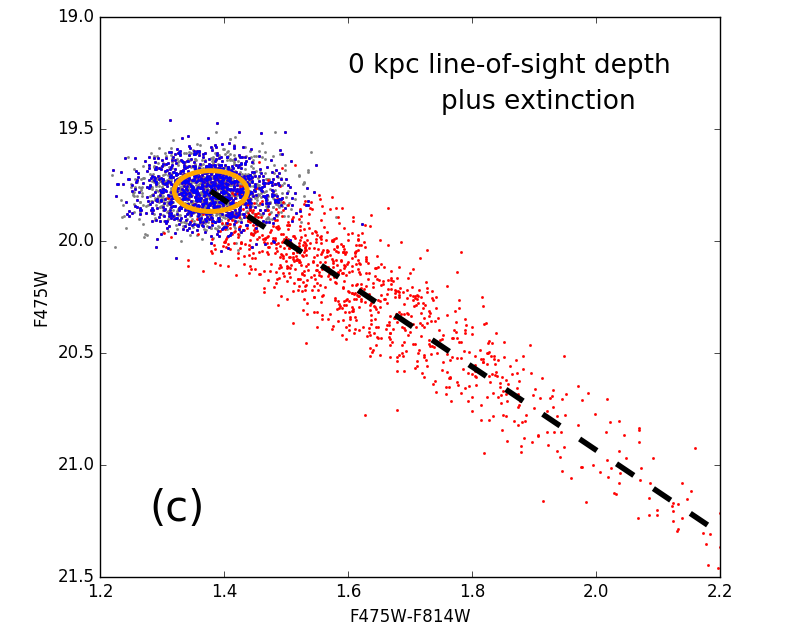}
\includegraphics[width=0.49\textwidth]
{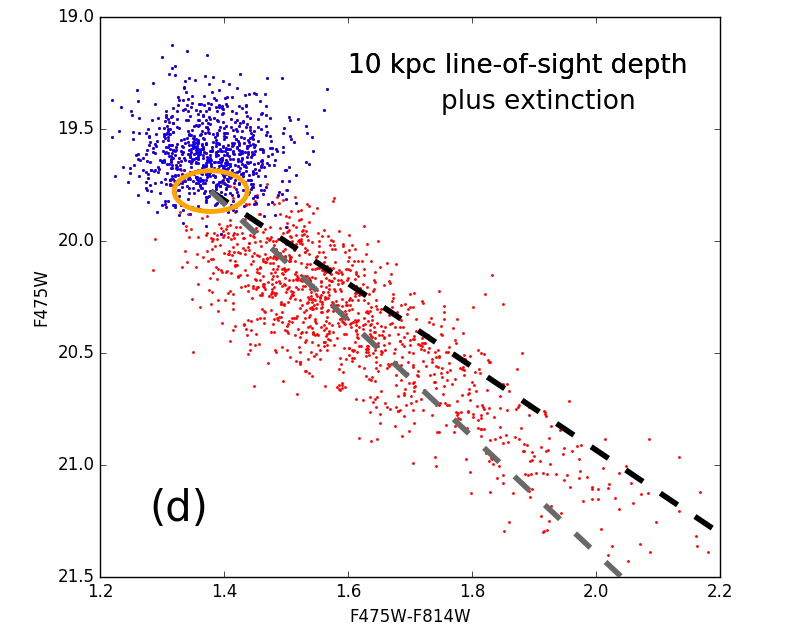}
\centering	\includegraphics[width=0.49\textwidth]
{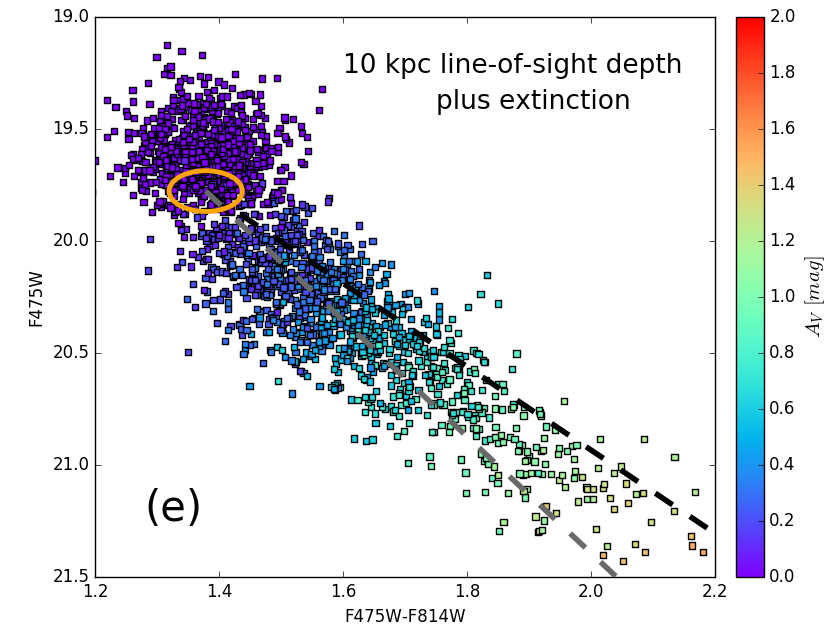}
\includegraphics[width=0.49\textwidth]
{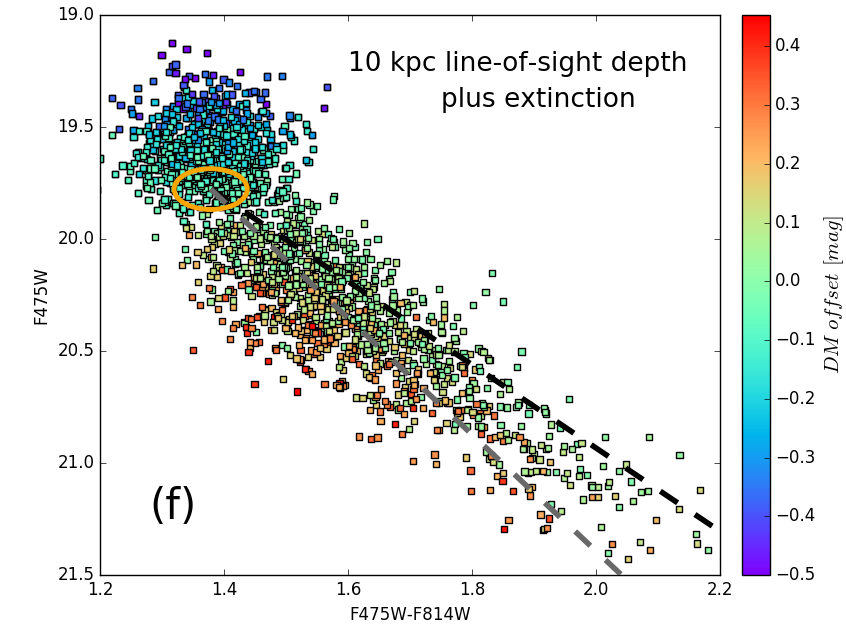}

\caption{ Red clump model simulating the line of sight depth effect. Each panel has 2000 red clump stars. Panels (a) and (b) show a simulated RC population with a mean distance of $\langle{D}\rangle$ = 62 kpc ($\mu_{0} = 18.96$ mag), unaffected by dust extinction and subject to two different SMC line-of-sight depths of zero kpc and 10 kpc to illustrate the impact of galactic structure on the RC. Due to the highly-elongated shape of the galaxy along the line of sight (as illustrated in Fig. \ref{fig:distributionLOSdepth}), stars have varying distances seen as a spread in magnitude in panel (b) where stars closer along the line of sight are in blue and stars farther along the line of sight are in gray. The orange ellipse indicates the unreddened location of the red clump with values listed in Table \ref{tab:unredRCloc}. Panels (c) and (d) show models with SMC line-of-sight depth of 0 and 10 kpc respectively where stars unaffected by dust remaining in the foreground are in blue and stars reddened due to dust are in red. The clump undergoes reddening with the following parameters: mean extinction in the V band of $\langle{A_{V}}\rangle$ = 0.4, width of the log-normal distribution of extinction $\langle{\sigma_{A_{V}}}\rangle$ = 0.3, a dust distance $D_{dust}$ = 60 kpc for the 10 kpc line-of-sight depth model, and a 0.65 fraction of reddened stars. Panels (e) and (d) show models with 10 kpc line-of-sight depth where each star is color-coded by $A_{V}$ and distance modulus offset. The slope of the input reddening vector, R$_{in}$, is indicated by the black dashed line in each plot. For comparison, the gray dashed line indicates the slope of the reddened red clump with a 10 kpc line-of-sight depth as recovered from the model. For the zero line-of-sight depth case the extinction vector follows R$_{in}$, but for the 10 kpc line-of-sight depth case, the vector is steeper since the reddened stars are further away and therefore fainter.}
   \label{fig:toymodel}
\end{figure*}

The reddened red clump is prominently visible in the SMIDGE CMDs. Figure~\ref{fig:cmdsAllBands} shows the CMDs of the spatially resolved stars in SMIDGE's eight \textit{HST} photometric bands using optical $\fsf-\eof$ color chosen for its best signal-to-noise ratio. In the absence of dust, the RC would be seen as a compact feature on the CMD at the red end of the horizontal branch and blueward of the red giant branch (RGB). Due to variable extinction by dust, however, SMIDGE observations reveal a RC which appears as a streak above and almost parallel to the main sequence (MS) with a tail extending towards redder colors and fainter magnitudes. Since the effects of dust are great compared to the photometric uncertainty of the RC, the feature can be easily studied. To measure the extinction and extract the extinction curve slope, we measure the slope of the reddened red clump streak in each CMD. The measurement requires carefully defining an unreddened RC color and magnitude which we discuss in the next section.

Another prominent feature in the SMIDGE CMDs is the bimodal red giant branch (RGB). The RGB would theoretically appear as a narrow and almost vertical sequence just redward of the RC. Due to dust, however, the RGB assumes a bimodal distribution and this bimodality allows us to infer that the dust is in a thin layer relative to the stars. If the dust and stars were well mixed, the stellar distribution would fully sample the extinction distribution and we would see a continuously reddened RGB, which is not what we observe. Assuming the RC and RGB stars follow the same spatial distribution, we can infer that same bimodality should be present for the RC\footnote{Note the unreddened RGB falls in the region between the unreddened and reddened red clump, making it difficult to observe this bimodality.} (i.e. a foreground unreddened population and a background reddened population). Since there are no regions in the SMIDGE field which are free from dust, we have no unreddened red clump reference from our data. To accurately measure the extinction curve shape, we therefore need to rely on a model (described in the rest of Sec. \ref{sec:cmdfitting}) for the RC and the reddening it experiences.

\subsection{A Model for the Unreddened Red Clump \label{sec:UnredRClocation}}

\begin{figure*} 
\centering
\includegraphics[width=\textwidth]
{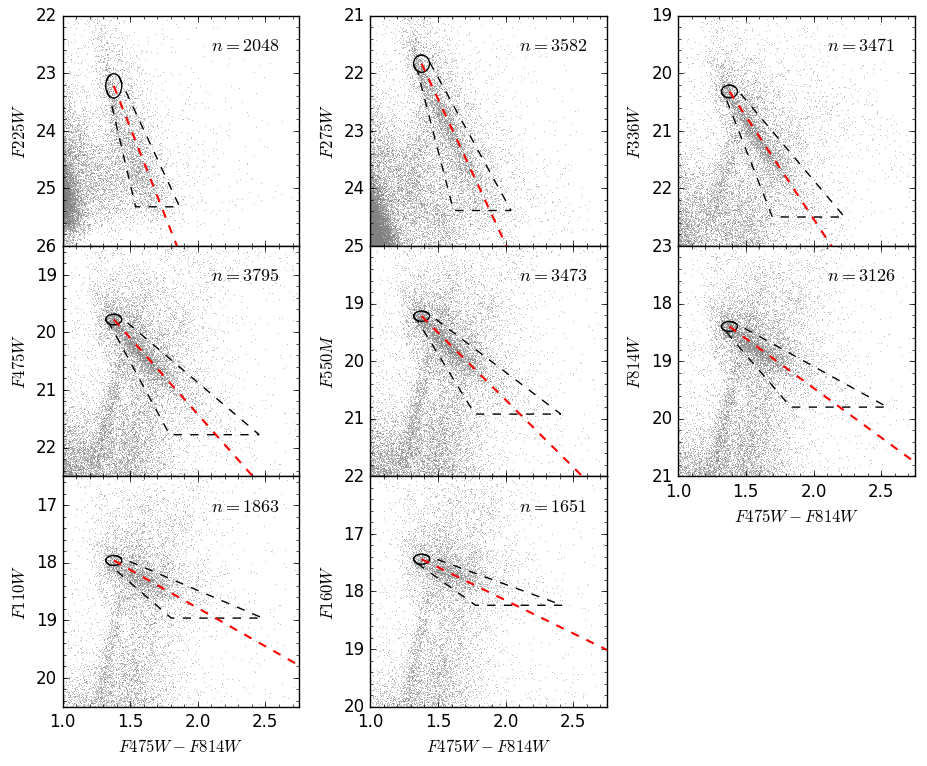}
\caption{Selecting reddened red clump stars from SMIDGE color-magnitude diagrams with a focus on the red clump. The cone-shaped selection region is shown bound by the black dashed line. The unreddened red clump appropriate to the Small Magellanic Cloud (see Section \ref{sec:UnredRClocation}) is indicated by the black ellipse. The size of the ellipse represents the intrinsic spread in color and magnitude of the theoretical red clump location due to the age range and metallicity of SMC red clump stars. The red dashed line represents the calculated reddening vector obtained as described in Section \ref{sec:SlopeMeasuring} with values given in Table \ref{tab:resultstable}. The size of the cone is determined by the approximate slope of the reddening vector with outer boundaries obtained by multiplying this slope by a factor of 1.4 in either direction (See Sections \ref{sec:SlopeMeasuring} and Appendix A). The number of sources within the selected reddened red clump region is indicated by $n$ for each photometric band.} 
\label{fig:cmdsPolySlope}
\end{figure*}

\begin{deluxetable}{lcc}
\tabletypesize{\scriptsize}
\tablecolumns{3}
\tablecaption{LMC Red Clump Properties}
\tablehead{ \multicolumn{1}{l}{ Filter} &
\multicolumn{1}{c}{Theoretical $RC_{Mag}$} &
\multicolumn{1}{c}{$\sigma_{Mag}$} }
\startdata
 \textit{F275W} &  22.20 & 0.12\\
 \textit{F336W} &  20.34 & 0.12\\
 \textit{F555W} &  19.16 & 0.08\\
 \textit{F775W} &  18.21 & 0.08\\
 \textit{F110W} &  17.72 & 0.10\\
 \textit{F160W} &  17.15 & 0.10
\enddata
\tablecomments{Unreddened LMC 30 Dor red clump location from \cite{GirardiSalaris:2001} and \citet[][DM16]{DeMarchi:2016} with a corresponding 1 $\sigma$ spread. MW foreground reddening of $E(B-V) = 0.07$ ($A_{V}=0.22$) has been applied. The adopted $F555W-F775W$ color for all photometric bands is 0.97$\pm$0.12}
\label{tab:unredRClocLMC}
\end{deluxetable}

\end{}

\begin{figure*}
  \centering
    \includegraphics[width=\textwidth]{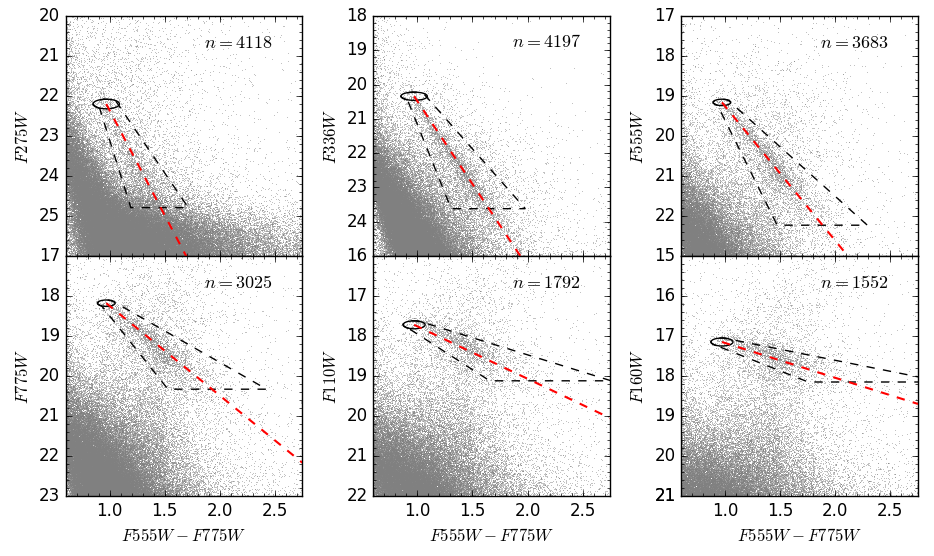}
    \vspace{0.2cm}
   \caption{ LMC 30 Dor CMDs from the HTTP survey with a focus on the red clump. See Figure \ref{fig:cmdsPolySlope} for a detailed description of the features on the CMDs. }
   \label{fig:httpcmds}
\end{figure*}

We construct a  model for the red clump as a two-dimensional Gaussian distribution of stars in color and magnitude. The key parameters which will set the model unreddened red clump (zero-point and widths in color and magnitude) are: age, metallicity and star formation history (SFH), average distance modulus, foreground reddening and SMC depth along the line of sight.  We define the red clump unreddened zero point and width by using a SFH incorporating the full ranges of ages and metallicities appropriate for the SMIDGE region.  This is an improvement upon previous works using this method which use single age and metallicity (e.g. \citetalias{DeMarchi:2014fe}, \citetalias{DeMarchi:2014en}, and \citetalias{DeMarchi:2016}).  We create a simulated CMD with foreground Milky Way extinction of A$_{V}=0.18$ \citep[derived from the Milky Way {\sc HI} foreground towards the SMIDGE field;][]{Muller:2003, Welty:2012} and R$_{V}=3.1$, zero line-of-sight depth, a distance modulus of 18.96 \citep{Scowcroft:2016hm}, and a SFH based on \cite{Weisz:2013mc}, shown in Figure \ref{fig:synthcmd}.  \cite{Weisz:2013mc} modeled the CMD of a nearby region in the SMC with similar RGB stellar surface density and with low internal dust attenuation based on deep \textit{HST} photometry. Older populations dominating the RC are well mixed spatially and we expect no significant difference between the \cite{Weisz:2013mc} field and SMIDGE, except for young ages (<200 Myr).  We use a simplified approximation of the SFH and age-metallicity relation (AMR) where we adopt a smooth star formation rate (SFR(t)), with SFR enhancements at 500 Myr and 1.5-3.0 Gyr and similar Z(t) but offset to lower values by 0.2-0.3 dex.  We compare the SMC SFH and AMR of \cite{Rubele:2015vmc} and \cite{Weisz:2013mc} and find good consistency between the RC properties obtained from our model and from the \cite{Rubele:2015vmc} results (where higher metallicity, but lower $A_{V}$ and older dominant RC ages compensate). We use the \texttt{fake} CMD simulation tool \citep[part of the \texttt{MATCH} software package;][]{Dolphin:2002} to produce the unreddened RC model, adopting the PARSEC stellar evolution models \citep{Bressan:2012, Tang:2014, Chen:2015} and a \cite{Kroupa:2001} initial mass function. The match between the observed location of the unreddened RGB in SMIDGE and the synthetic RGB give us confidence in the modeled CMD.  The above model allows us to account for a potentially complicated red clump morphology such as, for example, a secondary red clump composed of younger stars extending toward fainter magnitudes.

We create synthetic CMDs for all $\fsf-\eof$ color-magnitude combinations.  The resulting $\fsf$ CMD is plotted in Figure \ref{fig:synthcmd}, and the rest of the color-magnitude combinations are plotted in Figure \ref{fig:synthcmds} in the Appendix.  We model the position of the unreddened red clump using the full stellar population in the area, which includes red clump stars with a mean age of 1.81 $\pm$ 0.95 Gyr and a mean metallicity [M/H] = $-$ 0.95 $\pm$ 0.14. We select red clump stars in $\fsf-\eof$ matched across all eight CMDs by defining a wide box around the overdensity blueward of the red giant branch from which we pick objects which fall within a density contour encompassing 70\% of the total number of sources in the box. This subsample of stars has a mean age of 2.02 $\pm$ 0.66 Gyr and an average [M/H] = $-$ 0.90 $\pm$ 0.11. We use the 70 percentile contour to define a mean and a standard deviation for the red clump which is represented by the yellow ellipse in Figure \ref{fig:synthcmd}. The ellipse center is defined by measuring the centroid of the color and magnitude in the subsample in each model CMD where the width and height of the ellipse indicate the standard deviation in color and magnitude respectively due to the intrinsic age and metallicity spread of the RC (Table \ref{tab:unredRCloc}). This ellipse is then used to designate the theoretical unreddened red clump location in the observed CMDs.  In the Appendix we examine the effects of moderate variations in the zero point and in the width of the unreddened model RC.  Effects from line-of-sight binaries are negligible due to the brightness of the RC giant primary stars.  We additionally use artificial star tests to determine that the photometric error is small and therefore also negligible.

The theoretical location of the unreddened red clump population is dependent on the distance to the galaxy. We adopt an average SMC distance modulus of $\mu_{0}$ = 18.96 mag (62 kpc) from \cite{Scowcroft:2016hm}.  We test the sensitivity of our results to the assumed distance modulus in the Appendix since there is a well-known systematic distance variation as a function of position across the SMC.  This variation is demonstrated by observations of Cepheid and RR Lyrae variables, supergiants, red clump stars, or a combination of these populations \citep{Florsch:1981, CaldwellCoulson:1986, Welch:1987, HatzidimHawkins:1989, Gardiner:1991, GardinerHatzidimitriou:1992, Subramanian:2009, Subramanian:2012, Haschke:2012smc, Kapakos:2012, Nidever:2013ji, Scowcroft:2016hm, JD:2016a, JD:2016b, Subramanian:2017}. Several of these studies conclude that the galaxy is elongated along an axis approximately along our line of sight with a 10-kpc full-width-half-maximum (FWHM) of the stellar density distribution \citep{Gardiner:1991, GardinerHatzidimitriou:1992, Nidever:2013ji, JD:2016a, Subramanian:2017}. Since we extend our analysis to the Large Magellanic Cloud, we note that an LMC line-of-sight depth would similarly impact extinction curve results.  For the LMC \citet[][hereafter JD16]{JD:2016a} conclude a FWHM distribution along the line of sight of about 5 kpc, where the LMC distance is \about 50 kpc ($\mu_{0} = 18.48 \pm 0.04$ mag) \citep{Monson:2012}.  For both the Small and the Large Magellanic Clouds this means that stars will have distances which vary significantly relative to the distance of the center of the galaxy itself as illustrated in the left panel of Fig. \ref{fig:distributionLOSdepth}. This effect will inevitably be reflected on a color-magnitude diagram by creating a spread in the magnitude of stars.


\subsection{Modeling the Reddened Red Clump} \label{sec:ModelDust}

The line of sight depth of the SMC and the extinction the stars experience from the thin dust layer lead to an important effect: the unreddened red clump stars are closer (and therefore brighter) than the reddened red clump stars.  An illustration of the model is given in the right panel of Figure \ref{fig:distributionLOSdepth}.  To account for all contributions to the extinction curve shape, we show the combined effect of line-of-sight depth and extinction by including a thin dust layer in our model of the red clump.  We begin with a population of red clump stars positioned at the theoretical CMD red clump location described in Sec. \ref{sec:UnredRClocation}. We then assume the stars have a Gaussian distance distribution with a mean of 62 kpc average SMC distance \citep{Scowcroft:2016hm}.  In generating the model reddened red clump, we convolve the initial Gaussian distribution of magnitudes for the stars with this assumed Gaussian distance distribution.  As discussed in Sec. \ref{sec:RCextOverview}, we see a clear separation between the reddened and the unreddened red giant branch, and we can thus assume that we can neglect the dust layer's depth along the line of sight since it is negligible compared to the depth of the distribution of the stars. The dust in our model is thus positioned in front of a fraction of the red clump population which we set at 0.65.  Since this reddened fraction varies with the position of the dust relative to the stars, it is ultimately a function of the geometry of the region.  Varying the fraction between 0.65 and 0.35 steepens the red clump slope by 0.1 - 0.2 across photometric bands ranging from F160W to F225W respectively.  In a future study we plan to explore the way in which the relative geometry of the dust and stars affects the full synthetic CMD rather than only this simplified model of the red clump.

The model assumes the dust extinction experienced by the stars behind the screen has a log-normal column density distribution of $A_{V}$ sampled randomly by the stars. There are several reasons for choosing a log-normal distribution \cite[see][]{Dalcanton:2015bl}. Observations and simulations suggest that log normal probability density functions should be a ubiquitous feature of the turbulent interstellar medium \citep{Hennebelle:2012,Kainulainen:2009,Hill:2008}.

Our model employs the following parameters: $\langle{D}\rangle$ is the mean distance to the galaxy, $D_{FWHM}$ is the FWHM line of sight depth of the galaxy, the dust layer is located at a distance $D_{dust}$, the stars experience a mean extinction $\langle{A_{V}}\rangle$ where the width of the log-normal distribution of extinctions is $\langle{\sigma_{A_{V}}}\rangle$. The input extinction curve slope which corresponds to the slope of the reddening vector is $R_{in}$.

To illustrate the effect of the line of sight depth on the measured extinction curve, we generate synthetic red clump datasets using this model with an input extinction curve slope and a line-of-sight depth as in Figure~\ref{fig:toymodel}.  It is clear in panels c) and d) in the figure that the line of sight depth causes a steepening of the reddening vector.  This results from the fact that the reddened stars are more distant and therefore fainter than the unreddened stars. Panels (e) and (f) of Figure \ref{fig:toymodel} show the distance modulus offset and $A_{V}$ for each of the stars in the 10-kpc line-of-sight depth model depicted in panel (d).  The result of the Gaussian distribution of distance moduli and the log-normal distribution of $A_{V}$ is that for even the highest $A_{V}$ stars there is a range of distance moduli which, due to the Gaussian nature of the distribution, is weighted towards smaller offsets.  The slope of the reddened RC is determined by the mean of this combined distribution.
We also note that the theoretical, zero-depth red clump ellipse is at fainter magnitudes than the observed unreddened red clump. The reason for this is that the unreddened foreground RC population is closer. \cite{Nidever:2013ji}, \cite{Girardi:2016}, and \cite{Subramanian:2017} suggest the effect of an extended RC seen in color-magnitude diagrams can be attributed to the large line-of-sight depth rather than to population effects.

\begin{figure*} 
	\centering
    \includegraphics[width=0.493\textwidth]
{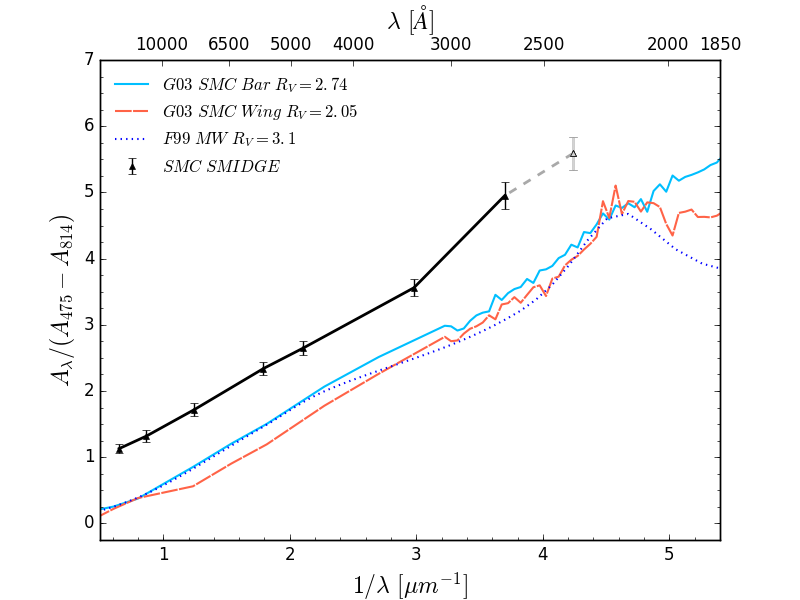}
\hspace{0.1cm}
	 \includegraphics[width=0.493\textwidth]
{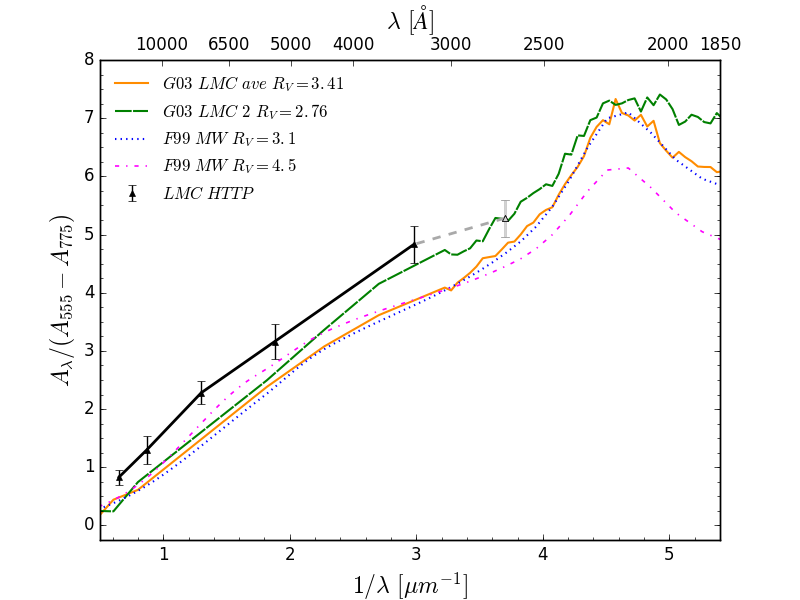}
\caption{ Left: The SMC SMIDGE extinction curve is plotted in black. For comparison, the \citet[][G03]{Gordon:2003} SMC Bar and SMC Wing curves, and the \citet[][F99]{Fitzpatrick:1999} Milky Way $R_{V}=3.1$ curves normalized to SMIDGE wavelengths are shown as well. On the right is the 30 Dor LMC extinction curve. The LMC average and LMC 2 \citetalias{Gordon:2003} curves, and the \citetalias{Fitzpatrick:1999} Milky Way $R_{V}=3.1$ and $\bm{R_{V}=4.5}$ curves are plotted as well. We conclude that the observed offsets in both the SMC and the LMC curves (the latter also noted by \citet[][DM16]{DeMarchi:2016}) are a result of the significant depth along the line of sight of both galaxies (see Sec. \ref{sec:UnredRClocation} ). }
   \label{fig:SmidgeHttpResultsMWcurves}
\end{figure*}

\subsection{Measuring the Red Clump Reddening Vector Slope}
\label{sec:SlopeMeasuring}

To extract the extinction curve shape from the observed color-magnitude diagrams we need to determine the slope of the vector along which the red clump is extended from its theoretical unreddened location. Figure \ref{fig:cmdsPolySlope} illustrates our technique.  To find the slope of the reddening vector using the method described below we need to first select the stars which will be used in the calculation since the extended red clump is not completely isolated on the CMDs but blends into other features. We select red clump stars by placing a boundary around these sources. First, we define a generous selection box surrounding the reddened red clump streak. We then find the approximate slope of the reddened red clump by fitting a linear slope to the points inside this region using the bisector of two lines found with the ordinary least squares method \citep{Isobe:1990}. To refine this value, we then narrow the selection region by defining upper and lower boundaries set by the width of the ellipse representing the unreddened red clump and tangents to this ellipse whose slope is plus or minus 40\% of the approximate slope. The bottom bound is set where the number density of sources in a region spanning one tenth of a magnitude falls below 20 sources. We have investigated the effects of these boundary choices on the output slope by performing a series of sensitivity tests as described in the Appendix.

To determine the reddening, which is canonically given by $E(B-V) = A_{B}-A_{V}$, we choose \textit{HST's} optical $\fsf$ and $\eof$ to obtain $E(\fsf-\eof)$ resulting in $R_{\lambda} = A_{\lambda}/(A_{475}-A_{814})$. $\fsf$ and $\eof$ have effective wavelengths of 474.7 nm and 802.4 nm and are closely related to the SDSS $g'$ and Johnson-Cousins $I$ filters with effective wavelengths of 477 nm and 806 nm respectively \citep{Dressel:2014}. We thus approximately obtain $R_{gI}$ = $A_{\lambda}$ / E(g$\prime$ $-I)$.

The boundaries are determined independently for each combination of magnitude and $\fsf-\eof$ color. Although some of the reddened sources may come from the RGB instead of strictly from the RC (see Figure \ref{fig:synthcmd}), since they are reddened due to dust along the same vector as RC stars are, they do not interfere with our slope measurement. We tested alternative ways to set the bounds of the RC selection, such as defining a wider or a narrower region around the apparent RC feature, extending the top and bottom bounds, or shortening the bottom bound. These variations and the results they produce are explored further in the Appendix, but they do not alter the results in addition to the reported uncertainties.

To measure the slope of the reddening vector using the refined selection of stars, we again use the bisector of the two ordinary least squares lines. The result is the ratio of the extinction in magnitudes (absolute extinction) to that in color (selective extinction), or $R_{\lambda}$. The values of this vector for the range of wavelengths explored by SMIDGE produce the extinction curve itself.

Uncertainties on calculating $R_{\lambda}$ using our method result from a combination of factors. One source of error results from the intrinsic spread of the unreddened red clump population caused by the range of ages and metallicities attributed to the clump.  Another is due to the scatter in the red clump distribution affecting the fit of the ordinary least squares bisector line to the red clump reddening vector.  Since the RC becomes almost vertical in $F225W$'s CMD, it becomes difficult to measure the slope of the reddening vector in this filter resulting in another source of error.  All these are taken into account and are used to report the uncertainty in our results. The photometric error for these bright sources as determined using artificial star tests is small and is not taken into account. An additional systematic uncertainty in $R_{225}$ which is not quantified results from \textit{HST's F225W} red leak.\footnote{The red leak affects the measured stellar magnitudes in \textit{F225W} by contributing to an off-band flux in the red part of the spectrum \citep{Dressel:2014}.}

\begin{deluxetable} {lccc}
\tablewidth{0pt}
\tabletypesize{\footnotesize}
\tablecolumns{4}
\tablecaption{SMC SMIDGE Extinction Curve Results}
\tablehead{ \multicolumn{1}{l}{$\lambda$  [\AA]} &
\multicolumn{1}{c}{$\lambda \textsuperscript{-1}$ [$\mu$m]} &
\multicolumn{1}{c}{Band Combination} &
\multicolumn{1}{c}{$R_{\lambda}$} }
\startdata
 2359 & 4.24 & $A_{225}/(A_{475}-A_{814})$ & 5.59 $\pm$ 0.23 \\
 2704 & 3.70 & $A_{275}/(A_{475}-A_{814})$ & 4.95 $\pm$ 0.19 \\
 3355 & 2.98 & $A_{336}/(A_{475}-A_{814})$ & 3.56 $\pm$ 0.13 \\
 4747 & 2.10 & $A_{475}/(A_{475}-A_{814})$ & 2.65 $\pm$ 0.11 \\
 5581 & 1.79 & $A_{550}/(A_{475}-A_{814})$ & 2.34 $\pm$ 0.10 \\
 8024 & 1.25 & $A_{814}/(A_{475}-A_{814})$ & 1.72 $\pm$ 0.09 \\
 11534 & 0.87 & $A_{110}/(A_{475}-A_{814})$ & 1.32 $\pm$ 0.08 \\
 15369 & 0.65 & $A_{160}/(A_{475}-A_{814})$ & 1.13 $\pm$ 0.07 
\enddata

\label{tab:resultstable}
\end{deluxetable}

\begin{deluxetable} {lcccc}
\tablewidth{0pt}
\tabletypesize{\footnotesize}
\tablecolumns{5}
\tablecaption{LMC 30 Doradus Extinction Curve Results}
\tablehead{ \multicolumn{1}{l}{$\lambda$  [\AA]} &
\multicolumn{1}{c}{$\lambda \textsuperscript{-1}$ [$\mu$m]} &
\multicolumn{1}{c}{Band Combination} &
\multicolumn{1}{c}{$R_{\lambda}$} &
\multicolumn{1}{c}{$R_{\lambda}  DM16\textsuperscript{1}$}}
\startdata
 2704 & 3.70 & $A_{275}/(A_{555}-A_{775})$ & 5.28 $\pm$ 0.32 & 5.15 $\pm$ 0.38\\
 3355 & 2.98 & $A_{336}/(A_{555}-A_{775})$ & 4.83 $\pm$ 0.32 & 4.79 $\pm$ 0.19\\
 5308 & 1.88 & $A_{555}/(A_{555}-A_{775})$ & 3.16 $\pm$ 0.30 & 3.35 $\pm$ 0.15\\
 7647 & 1.31 & $A_{775}/(A_{555}-A_{775})$ & 2.28 $\pm$ 0.20 & 2.26 $\pm$ 0.14\\
 11534 & 0.87 & $A_{110}/(A_{555}-A_{775})$ & 1.30 $\pm$ 0.24 & 1.41 $\pm$ 0.15\\
 15369 & 0.65 & $A_{160}/(A_{555}-A_{775})$ & 0.83 $\pm$ 0.13 & 0.95 $\pm$ 0.18
\enddata
\tablenotetext{1}{Note that \citetalias{DeMarchi:2016} call this value R and not $R_{\lambda}$.}
\label{tab:resultstableLMC}
\end{deluxetable}

\begin{figure*} 
  \centering 
    \includegraphics[width=0.4935\textwidth]{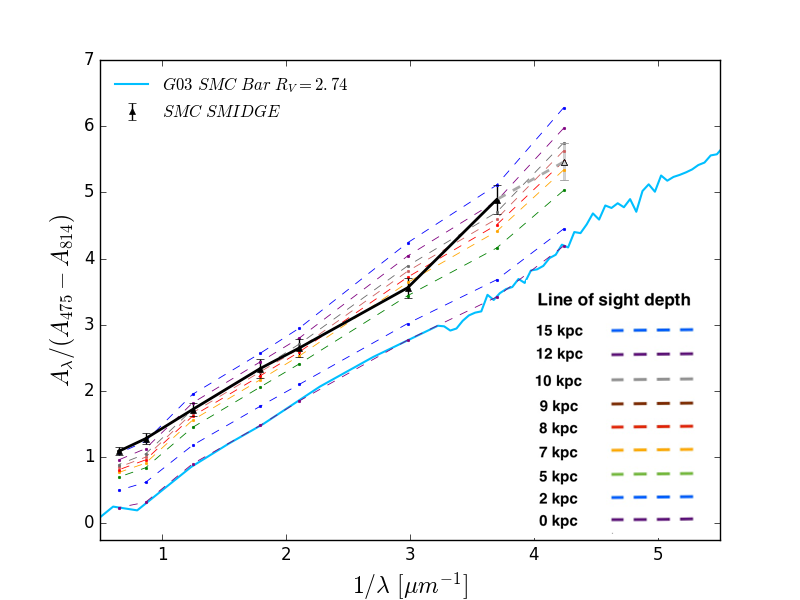}
    \hspace{0.1cm}
    \includegraphics[width=0.4935\textwidth]{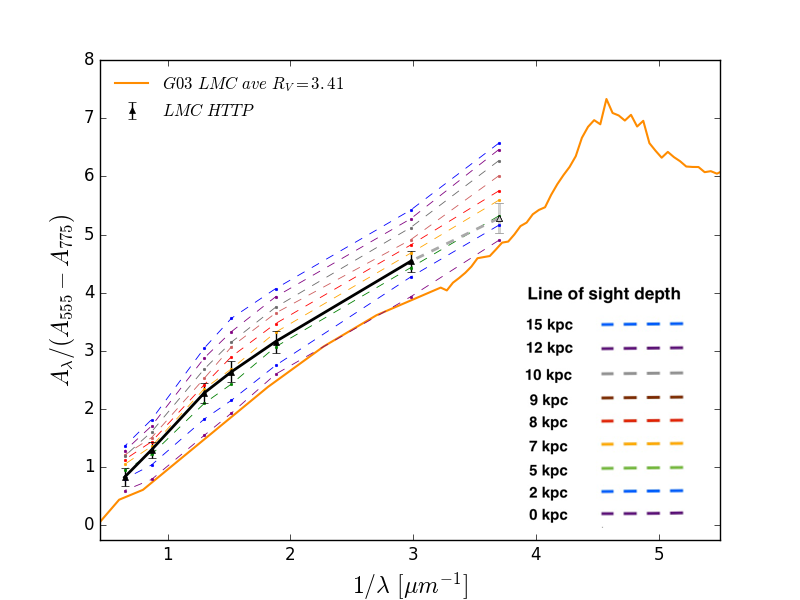}
    \includegraphics[width=0.4935\textwidth]{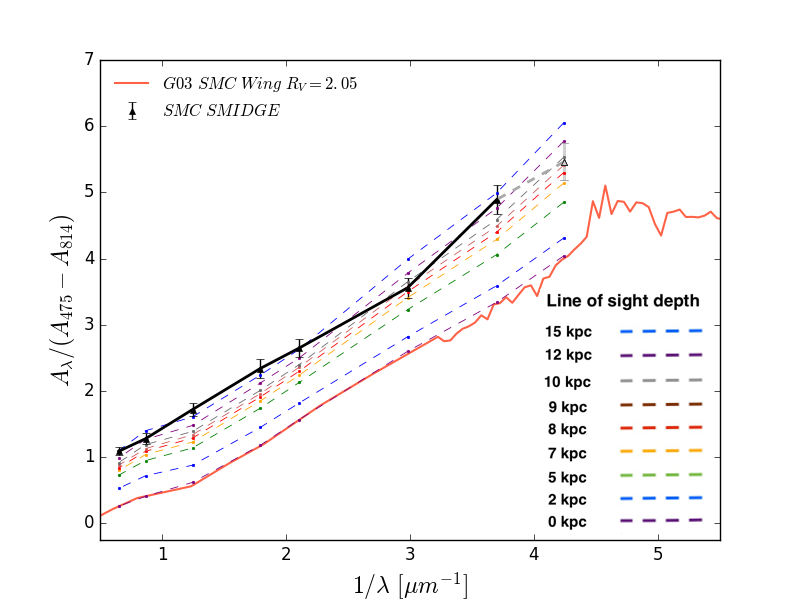}
    \includegraphics[width=0.4935\textwidth]{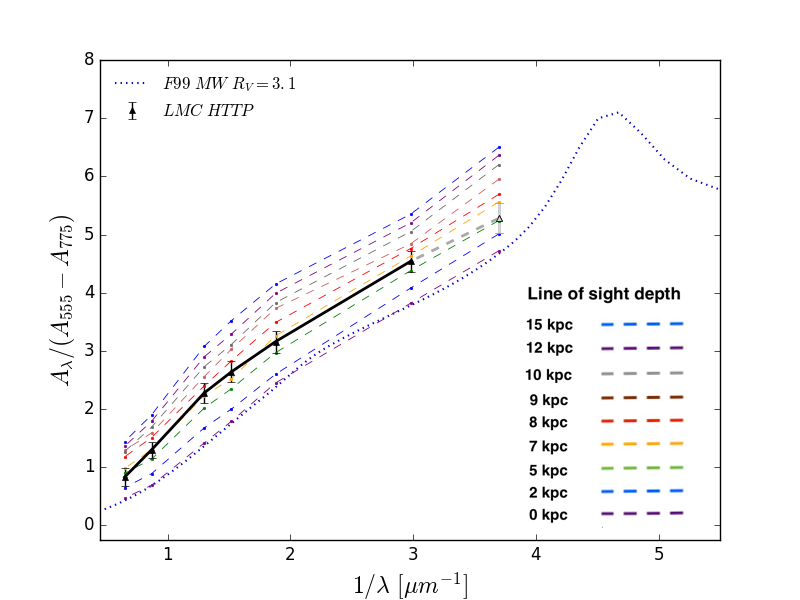}
   \caption{ Effect of the line of sight depth of the SMC and the LMC on the SMC and LMC extinction curves of \citetalias{Gordon:2003} and the MW $R_{V}$=3.1 curve of \citetalias{Fitzpatrick:1999} showing the offset in the presence of galactic depth between 0 and 15 kpc. The SMC SW Bar extinction curve found in this study and plotted in black in the two panels on the left indicates a line-of-sight depth of 10 $\pm$ 2 kpc when compared to the \citetalias{Gordon:2003} SMC Bar curve, and a 12 $\pm$ 2 kpc when compared to the G03 SMC Wing curve.  The LMC 30 Doradus extinction curve is plotted in black in the panels on the right, and indicates it is probing a region with a line of sight depth of 5 $\pm$ 1 kpc when compared to the \citetalias{Gordon:2003} LMC average curve and 7 $\pm$ 1 kpc when compared to the \citetalias{Fitzpatrick:1999} MW $R_{V}$=3.1 curve. Details of this analysis are in Sec. \ref{sec:LOSdModelExtCurves}. }  \label{fig:resultsAndModeledLOSdepthEffect}
\end{figure*}

We perform the same analysis on data from the 30 Doradus region in the Large Magellanic Cloud from the Hubble Tarantula Treasury Project \citep{Sabbi:2013,Sabbi:2016} to compare to the \citetalias{DeMarchi:2016} extinction curve results. The HTTP survey uses observations in near-ultraviolet, optical, and near-infrared wavelengths in the range 0.27-1.5 $\mu$m. The unreddened locations for red clump stars in the 30 Dor region are those used by \citetalias{DeMarchi:2014fe} based on \citet{GirardiSalaris:2001} models which \citetalias{DeMarchi:2014fe} identify (in Section 3.1 of their paper) as the RC stars with the lowest metallicity which fit the observations. To define the magnitude of the RC they conclude a metallicity of Z = 0.004 for the oldest (> 1 Gyr) RC stars in the 30 Dor region. To narrow down the average color for the RC they select stars of ages 1.4-3.0 Gyr. These unreddened RC location values are listed in Table \ref{tab:unredRClocLMC}.  The extended red clump feature in the LMC color-magnitude diagrams is defined in the same way as the red clump in the SMIDGE analysis. 30 Dor color-magnitude diagrams are shown in Figure \ref{fig:httpcmds}.  Since the LMC is not as deep along the line of sight as the SMC, there is less vertical spread in magnitude which in turn allows for a clearer definition of the unreddened red clump in the LMC CMDs in Fig. \ref{fig:httpcmds}.  At the same time the LMC stars studied here are much more heavily reddened than the SMC stars and this results in an extended and noticeable reddened red clump streak.  LMC uncertainties in reddening vector calculations stem from similar sources as those in the SMIDGE analysis discussed above.  In our final analysis we calculate the slope for six LMC photometric bands - \textit{F275W, F336W, F555W, F775W, F775U, F110W}, and \textit{F160W}.

\subsection{Results}
\label{sec:pre-results}

$R_{\lambda}$ results for the SMC and the LMC are given in Tables \ref{tab:resultstable} and \ref{tab:resultstableLMC}.  We plot the extinction curves in Figure \ref{fig:SmidgeHttpResultsMWcurves} along with other known extinction curves for comparison, such as those of \citetalias{Gordon:2003} and \citet[F99 hereafter]{Fitzpatrick:1999}. Our results for the LMC are in good agreement with the results of \citetalias{DeMarchi:2016}. For both the SMC and LMC we observe a larger $R_{\lambda}$ than indicated by existing spectroscopic extinction curve measurements.  For comparison, the new SMC extinction curve has $R_{475} = A_{475}/(A_{475}-A_{814}) = 2.65 \pm 0.11$, while the \citetalias{Gordon:2003} equivalents are $R_{475}^{SMCbar}$ = 1.86 and $R_{475}^{SMCwing}$ = 1.57, and the \citetalias{Fitzpatrick:1999} Milky Way $R_{V}$=3.1 equivalent is $R_{475}^{MW}$ = 1.83.  The LMC extinction curve we measure has $R_{555} = A_{555}/(A_{555}-A_{775}) = 3.16 \pm 0.3$, while the \citetalias{Gordon:2003} equivalents are $R_{555}^{LMCave}$ = 2.48 and $R_{555}^{LMC2}$ = 2.61, and the \citetalias{Fitzpatrick:1999} Milky Way $R_{V}$=3.1 equivalent is $R_{555}^{MW}$ = 2.37.  In the section which follows we discuss the interpretation of these measurements in light of recent observations that both the SMC and the LMC have a substantial line-of-sight depth which significantly impacts the calculations.

\section{Interpreting Extinction Results} 
\label{sec:extinctionresults}

\subsection{Line-of-Sight Depth Effect on Model Extinction Curves}
\label{sec:LOSdModelExtCurves}

We use our reddened red clump model to determine what the line of sight depth impact would be on SMC and LMC extinction curves observed by \citetalias{Gordon:2003} and modeled by \citet[WD01 hereafter]{WD:2001}, as well as the \citetalias{Fitzpatrick:1999} model Milky Way curves.  We begin by taking the $R_{\lambda}$ value of an observed or modeled extinction curve as an input extinction curve slope $R_{in}$ for the red clump model. The unreddened RC locations are those specified in Tables \ref{tab:unredRCloc} and \ref{tab:unredRClocLMC}. We then apply the procedure described in Section \ref{sec:ModelDust} while we vary the line of sight depth between 0 and 15 kpc to extract an output slope $R_{out}$. The results plotted in Figure \ref{fig:resultsAndModeledLOSdepthEffect} indicate that an increasing depth along the line of sight steepens the reddening vector slope such that the extinction curve produced as a result experiences an offset towards higher $R_{\lambda}$.

As shown by the monotonic increase in $R_{out}$ with depth, the line-of-sight depth effect is color-independent to first order in that all wavelengths are affected by the extended structure along the line of sight approximately equally. Essentially, the line of sight depth means that the reddened red clump stars are further away than the theoretical zero-point would suggest, leading to the reddened stars being offset to fainter magnitudes.  In essence, distance offsets are mimicking a truly ``gray'' extinction curve.  This fact allows us to separate the distance effect from the effect of extinction by dust. However, our sensitivity tests, described in the Appendix, illustrate that the assumed average RC distance modulus is somewhat degenerate with the line-of-sight depth shown in Fig. \ref{fig:resultsAndModeledLOSdepthEffect}. In the sensitivity tests we move the RC zero-points for all filter combinations by $\pm$0.15 mags, which in effect shifts the average distance of the RC stellar distribution by $\pm$4.5 kpc \citep[we note this is well outside the measured uncertainty of the average distance to the SMC;][]{Scowcroft:2016hm}.  The shift in $R_{out}$ introduced by this offset is color independent to first order, as is the line-of-sight depth effect. We find that a + $0.15$ mag shift in the RC zero-point would lead to a line-of-sight depth measurement of $\sim7$ kpc, rather than 10 kpc, and a - $0.15$ mag shift leads to an even larger line-of-sight depth. This test shows that for a reasonable range of SMC distance moduli, explaining our measured reddening vector slopes necessitates a large line of sight depth.

To test the degree of agreement between the resulting SMC SMIDGE region extinction curve and the extinction curves of \citetalias{Gordon:2003}, \citetalias{Fitzpatrick:1999}, and \citetalias{WD:2001}, and to extract the line of sight depth for the galaxy from RC stars, we find the minimum $\chi^2$ value by comparing our measured $R_{\lambda}$ values to those for modeled reddened red clumps with varying line-of-sight depths and input extinction curves. A similar measurement is performed for our LMC extinction curve results. Our results suggest that the observed SMC and LMC red clump extinction curves indicate a significant depth along the line of sight in both galaxies. For the SMC, we find the $\chi^2_{min}$ at a line-of-sight depth of 10 kpc $\pm$ 2 kpc at a 90 \% confidence level for the SMC Bar extinction curve. Comparing the SMIDGE extinction curve with other curves (\citetalias{Gordon:2003}'s SMC Bar ($R_{V} = 2.74$), SMC Wing (AzV 456, $R_{V} = 2.05$), LMC2 Supershell ($R_{V}=2.76$), LMC Average ($R_{V}=3.41$), \citetalias{Fitzpatrick:1999}'s MW $R_{V}=3.1$ and MW $R_{V}=5.5$, and with \citetalias{WD:2001}'s curves), we find that the newly-derived SMC extinction curve compares almost equally well with both the G03 SMC Bar and SMC Wing extinction curves (see left-side panels of Fig. \ref{fig:resultsAndModeledLOSdepthEffect}). To differentiate between the two, our future work will use UV bright stars to study the \textit{F225W} extinction, which will eliminate issues with the red leak. 

For the LMC 30 Dor region we obtain a $\chi^2_{min}$ at a line-of-sight depth of 5 kpc $\pm$ 1 kpc at the 90 \% confidence level. Comparing the LMC extinction curve with the above curves, we find that this curve similarly does not favor a single observed extinction curve but that it instead can be well-described by either the \citetalias{Gordon:2003} LMCave extinction curve ($R_{V}=3.41$) or the \citetalias{Fitzpatrick:1999} MW $R_{V}=3.1$ curve.  The latter would suggest a 7 $\pm$ 1 kpc line-of-sight depth.

\subsection{Comparison to Other Extinction Curve Shape Measurements} \label{sec:compareResults}

Extinction curves in the SMC and the LMC have been measured spectroscopically by \citet{Gordon:1998}, \citetalias{Gordon:2003}, \citet{MaizApellaniz:2012} and \citet{MA:2014}.  \citetalias{Gordon:2003} derive extinction curves for the SMC and the LMC via the pair method using ultraviolet spectroscopy and optical and near infrared photometry. They base their conclusions on five stars in the SMC, four of which are in the SMC Bar producing an average $R_{V}=2.74 \pm 0.13$, and one star in the SMC wing with $R_{V}=2.05 \pm 0.17$.  They also measure an average LMC2 Supershell ('LMC2') $R_{V}=2.76 \pm 0.09$ for nine stars and an LMC Average $R_{V}=3.41 \pm0.06$ for ten stars. These extinction curves are plotted in Fig. \ref{fig:SmidgeHttpResultsMWcurves} for comparison with the extinction curve results from this study.  We observe a general consistency in our results for the shape of the SMC and LMC extinction curves with the shape of the \citetalias{Gordon:2003} curves for wavelengths longward of \textit{F225W}'s.  At the same time we note the offset between the two sets of curves which we attribute to the effect of the depth of the two galaxies along the line of sight described in Sec. \ref{sec:UnredRClocation}.  We can not make a strong statement about the UV portion of the curve as we are limited by photometric effects when using red clump stars.

\citetalias{DeMarchi:2014fe}, \citetalias{DeMarchi:2014en}, and \citetalias{DeMarchi:2016} use UV-IR multiband \textit{HST} photometry to examine 30 Doradus in the LMC.  They find an offset between their results and the results of \citetalias{Gordon:2003} and the canonical Galactic extinction curves due to red clump reddening vector slopes which are considerably steeper than existing measurements.  Our measurements reproduce \citetalias{DeMarchi:2016}'s results. The authors attribute their results to the presence of ``gray'' extinction at optical wavelengths due to the vertical offset of their curves from the Galactic and \citetalias{Gordon:2003} curves.  Our model explains this effect as the result of the line-of-sight depth of the LMC where the extinction curve is well-reproduced by the \citetalias{Gordon:2003} LMC average $R_{V}$=3.41 extinction curve with a 5-kpc line-of-sight depth (see Fig. \ref{fig:resultsAndModeledLOSdepthEffect}).
 
\cite{MaizApellaniz:2012} use UV spectroscopy to obtain the extinction for four stars in the SMC quiescent cloud B1-1, and NIR/optical photometry to obtain the extinction curve for the five SMC stars studied by \citetalias{Gordon:2003}.  They conclude a significant variation from star to star in the extinction curve for the SMC B1-1 stars, particularly in the strength of the 2175 {\AA} bump.  Their results may imply that their sources are sampling ISM environments with a different dust composition.  In the section which follows we discuss the type of environment probed by the SMIDGE region and conclude that our results are most likely dominated by the diffuse ISM.  However, we currently do not have a conclusive result about the strength of the 2175 {\AA} bump due to the limitations of our red clump photometry.

\cite{MA:2014} use spectroscopy and NIR and optical photometry of O and B stars to derive the extinction curve inside 30 Doradus.  They find an $R_{V}$ equivalent which is also larger than the $R_{V}$ suggested by Galactic or \citetalias{Gordon:2003} extinction cures.  Although our results for 30 Dor agree with the authors' $R_{V}$=4.5 curve at optical wavelengths when the curves are expressed in $E(B-V)$ instead of $E(F555W-F775W)$ using spline interpolation, our extinction curve is better reproduced at NIR wavelengths by the \citetalias{Gordon:2003} LMC average $R_{V}$=3.41 curve with a 5-kpc line-of-sight depth.  The latter is also the case when the curves are normalized to $A_{V}$.  We obtain very similar results when we compare the \citetalias{DeMarchi:2016} curve (with values in the last column of Table \ref{tab:resultstableLMC}) to the \cite{MA:2014} curve.

Recent work by \cite{Hagen:2016smc} uses SMC UV, optical, and IR data integrated into 200'' regions to measure the attenuation curve of the SMC.  They find a 2175 {\AA} bump in most of the galaxy and a dust curve which is steeper than the Galactic curve.  We note that their study measures the attenuation curve (i.e. the combined effects of extinction, scattering and geometry (see \cite{Calzetti:1994})).  Our results, which are extinction curves, are not directly comparable to this study.

Red clump stars have been extensively used as a distance indicator to objects within the Milky Way and nearby galaxies \citep[][also see \cite{Girardi:2016} and references therein]{Cannon:1970, GirardiSalaris:2001, Bovy:2014}. Within the Galaxy in particular, they have been used to indicate the distance to the Galactic Center \citep{Paczynski:1998,Alves:2000,Francis:2014} and to open clusters \citep{Percival:2003}.  Distribution of stellar distances in the Magellanic Clouds have been measured with RR Lyrae and Cepheids \citep{Haschke:2012smc, Subramanian:2012, Scowcroft:2016hm, JD:2016a, JD:2016b}, and with red clump stars \citep{Subramanian:2017}. Our results in Sec. \ref{sec:LOSdModelExtCurves} indicating that the SMC has a depth of \about 10 kpc along the line of sight and that the LMC's depth is \about 5 kpc are generally consistent with distances derived from the Cepheids and RR Lyrae studies above.

\section{Discussion}
\label{sec:discussion}
We present measurements of the average extinction curve shape in the southwest bar region of the Small Magellanic Cloud covering an area of 100 $\times$ 200 pc using red clump stars. Our results indicate an extinction curve consistent with measurements from UV spectroscopy performed by \citetalias{Gordon:2003}. The significantly elongated structure of the SMC along the line of sight causes a perceived steepening of the reddening vector which if not accounted for would give the appearance of an extinction curve with a ``gray'' component. The latter is the conclusion of \citetalias{DeMarchi:2016} for the extinction curve of the 30 Doradus nebula in the LMC, which they attribute to the presence of an additional component due to gray dust. We conclude that gray extinction is not necessary to explain the observations in the SMC and the LMC when one accounts for the depth of the stellar distribution along the line of sight. 

There are several implications of our measurements. By sampling a relatively large region of the interstellar medium in the SMC, we can compare our averaged extinction curve results to measurements derived using the pair method targeting individual stars. We do this while noting that compared to the pair method, the technique we use presents a number of advantages, one of which is that by observing red clump stars on a color-magnitude diagram we do not need to know the location of the dust since we are taking an average measure of dust properties. One of the pair method's major limitations comes from having to carefully select stars in regions containing a significant amount of dust. At the same time, due to photometric limitations resulting from \textit{HST}'s red leak and the faintness of the stars in \textit{F225W}, the red clump technique we use in this paper does not allow us to have a strong handle on the strength of the 2175 Angstrom bump or the far UV rise.

One comparison we can make is with the work of \citetalias{Gordon:2003} who studied SMC extinction with the pair method. An interpretation of the similarity between our results and those of \citetalias{Gordon:2003} may be that over a relatively large area in the SMC there is little variation in the extinction curve shape.  Another possibility is that both studies are sampling the same ISM phases.  The dust probed with O- and B-type stars such as those from the UV spectroscopy sight lines of \citetalias{Gordon:2003} is generally assumed to be located in the diffuse envelopes of molecular clouds which would bias the extinction curves towards regions of star formation dominated by grain growth.  However, the \citetalias{Gordon:2003} SMC Bar $A_{V}$ extinctions of 0.35 - 0.68 indicate that these sight lines are probing the diffuse ISM, or at least regions where coagulation and grain growth are unlikely to have happened.

To address the hypothesis that both sets of extinction curves are probing the same ISM phases (e.g., diffuse atomic or molecular gas), we assess the type of environment the SMIDGE survey area covers by calculating the fraction of stars along lines of sight that have abundant molecular gas. Using APEX $^{12}$CO (2$-$1) mapping of the SW Bar at 28$''$ resolution (A. Bolatto, private communication), we make a conservative cut at I$_{CO}=$ 1 K km $s^{-1}$ to define molecular regions. We find that 23 \% of the red clump stars are found toward such regions, indicating that our new average extinction curves are dominated by diffuse material.  It remains a possibility that both \citetalias{Gordon:2003} and our work are sampling mainly the diffuse ISM, rather than molecular gas where grain growth may occur. In future work we will investigate the change of the extinction curve with ISM phase using the full stellar populations available from SMIDGE.

Another implication of our study is that extinction in the SMC and the LMC can be explained without the need to invoke ``gray'' extinction as suggested by \citetalias{DeMarchi:2014fe}, \citetalias{DeMarchi:2014en}, and \citetalias{DeMarchi:2016} for the LMC. If there were gray dust, the latter suggest that the reason may be the selective addition of fresh large grains due to Type II supernova explosions in 30 Dor. If recent supernova were the cause, this could suggest a bias towards gray extinction for galaxies with high star formation rates.  Gray dust also has implications for studies of the cosmological expansion with supernovae as their faintness could be interpreted as extinction without much reddening which would invariably impact inferring distances on cosmological scales.  Observationally there are demonstrations of the presence of gray extinction \citep{Strom:1971,Dunkin:1998,Gall:2014}.
However, our study concludes that extinction curves in the LMC can be explained without the need for an extra ``gray'' component of the dust, thus removing the argument for supernova dust production modifying the extinction curve.

\section{Conclusions}
\label{sec:conclusions}

We use color-magnitude diagrams based on SMIDGE \textit{HST} multiband photometry to measure the slope of the reddening vector of red clump stars in the southwest bar of the Small Magellanic Cloud in order to derive the extinction curve shape in the region. After noting that the depth along the line of sight of the SMC has a significant bearing on extinction curve shape results using this method, we model this effect to understand its impact on our results. When we properly account for the line-of-sight depth and analyze its effect on extinction curves, we conclude that the effect is significant and tends to give the appearance of steeper reddening vector slopes which in turn produce what appears to be gray extinction. Motivated by recent extinction curve shape results for 30 Dor in the Large Magellanic Cloud by \citetalias{DeMarchi:2014fe} and \citetalias{DeMarchi:2016} who use the same red clump method and also report an offset, we perform the same analysis on 30 Dor in the LMC.

Our conclusions for the optical and near-infrared portions of the extinction curve shape in both the SMC and the LMC is consistent with previous work such as the analysis of \citetalias{Gordon:2003} using spectroscopic measurements. Since depth effects in both galaxies produce an offset in the extinction curve due to a perceived steepening of the reddening vector, we conclude that it is the shape of the \citetalias{Gordon:2003} curves which our extinction curves match rather than a specific $R_{V}$ value.  Our recommendation to correct for dust extinction for individual objects is to therefore use the extinction curves derived by \citetalias{Gordon:2003}.  Correcting for dust extinction when using a full stellar population, however, calls for the need to account for the depth of the SMC and the LMC.  Additionally, we show that it does not have to be the case that gray extinction is responsible for the offset in the LMC 30 Dor extinction curve as \citetalias{DeMarchi:2014fe} and \citetalias{DeMarchi:2016} conclude, but that rather one needs to account for the depth along the line of sight when using methods relying on stellar distances to determine the extinction curve shape.  Future work aims at modeling the effect of extinction on all CMD features rather than simply on a generic red clump. Such an analysis will provide a way to understand the more subtle effects of dust on the extinction curve shape. \\
\\

\begin{figure} 
  \centering 
    \includegraphics[width=0.49\textwidth]{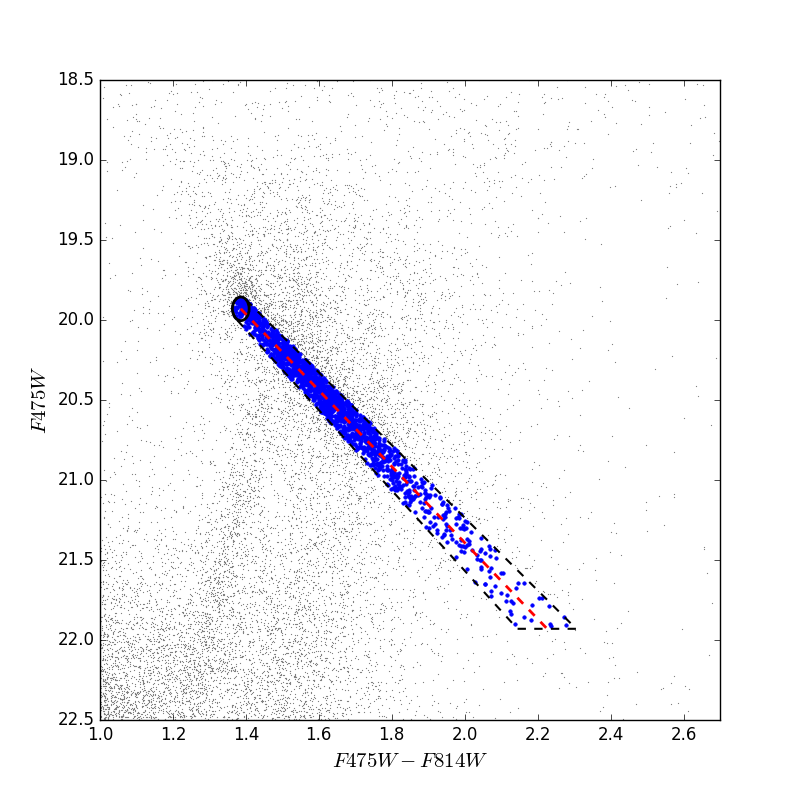}
    \hspace{0.1cm}
    \includegraphics[width=0.49\textwidth]{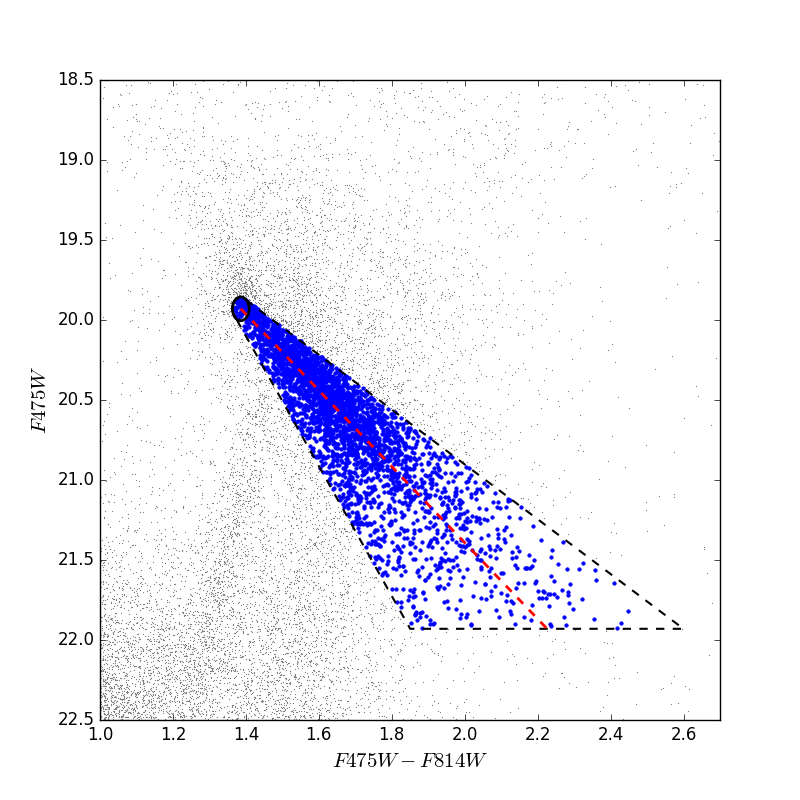}
   \caption{Red clump selection with variations in the approximate reddening vector slope to determine the effect of the selection on the slope calculation. The top panel shows a selection box corresponding to a \bm{$\pm$} 5\% variation in the approximate slope; the bottom panel shows the selection after a \bm{$\pm$} 40\% variation is added.  The blue points represent the highlighted reddened red clump stars inside the selection region. }
   \label{fig:vectorSlopeSens}
\end{figure}

\begin{figure} 
  \centering 
    \includegraphics[width=0.43\textwidth]{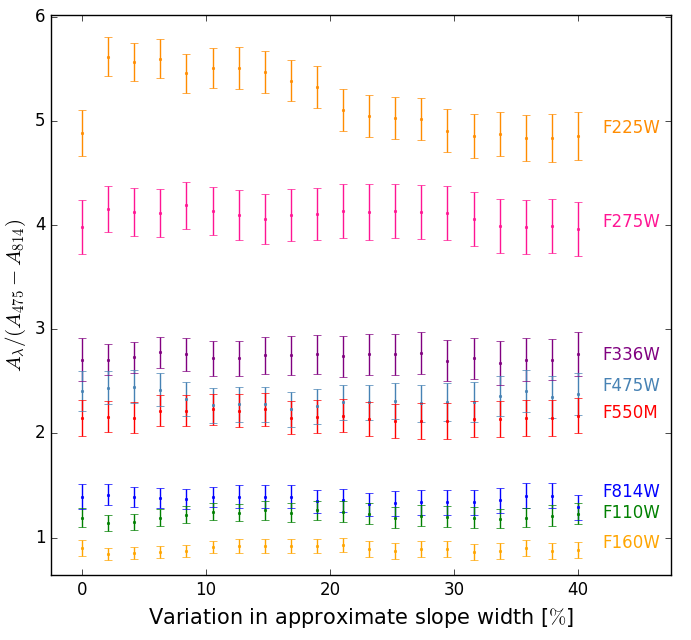}
    \hspace{0.025cm}
    \includegraphics[width=0.43\textwidth]{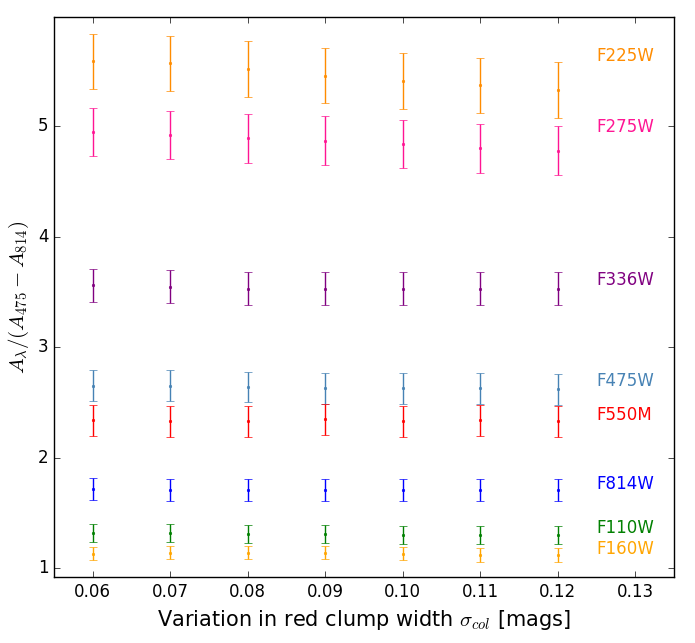}
\hspace{0.025cm}
    \includegraphics[width=0.43\textwidth]{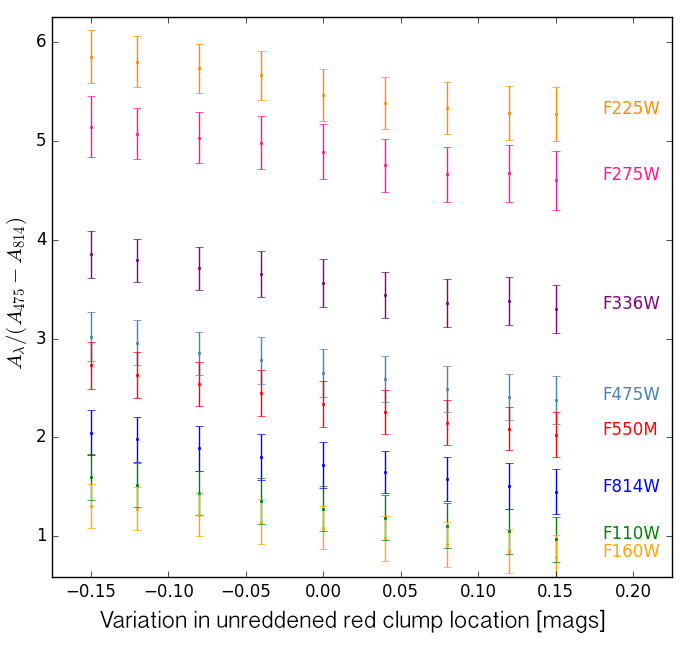}
\caption{Sensitivity of the calculated reddening vector slope to three parameters. The top plot shows the sensitivity to the approximate slope which determines the width of the reddened red clump selection region. The middle plot tests the sensitivity to the intrinsic width in color of the unreddened red clump. The bottom plot illustrates the sensitivity to the unreddened red clump location.}
   \label{fig:approxSlopeWidthSens}
\end{figure}

\begin{figure*}
  \centering
    \includegraphics[width=\textwidth]{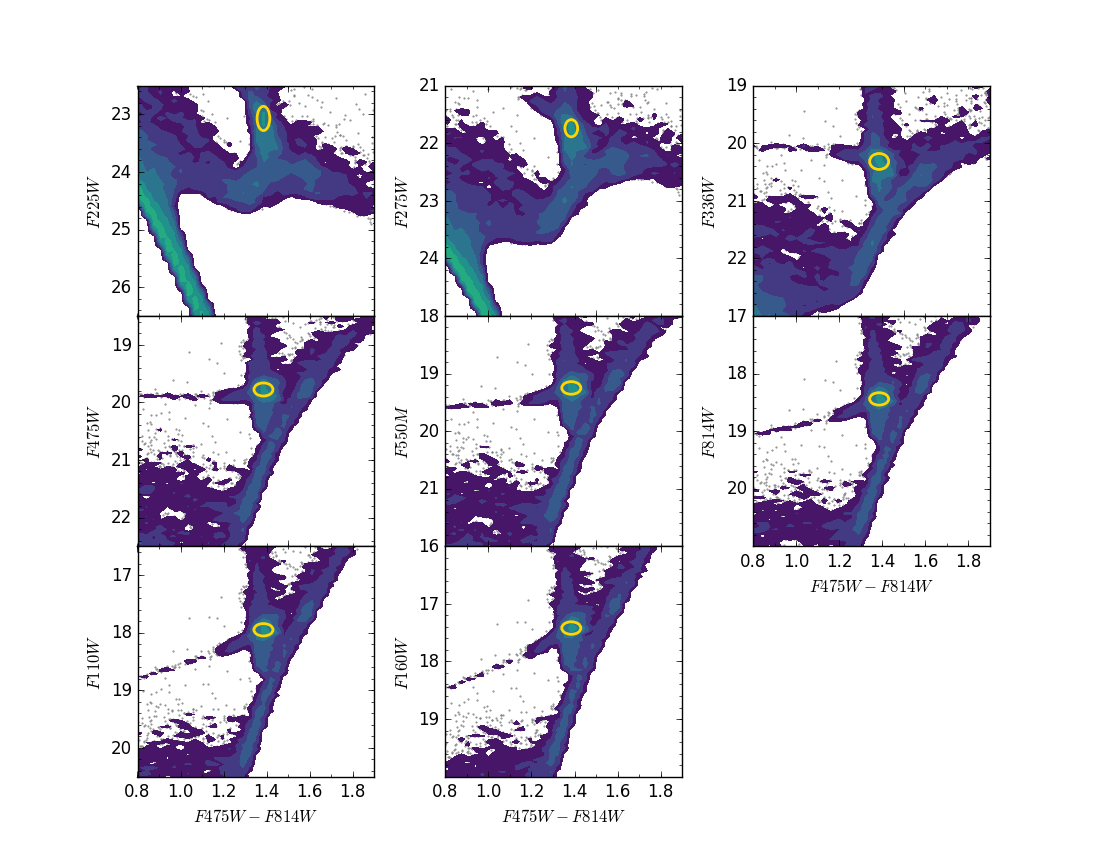}
    \vspace{0.2cm}
   \caption{ SMC synthetic CMDs generated to determine the location of the unreddened red clump (yellow ellipse, with values indicated in Table \ref{tab:unredRCloc}) as described in Section \ref{sec:UnredRClocation} and Figure \ref{fig:synthcmd}. }
   \label{fig:synthcmds}
\end{figure*}

We thank the referee whose comments helped improve this work. Support for this work was provided by NASA through grant number HST-GO-13659 from the Space Telescope Science Institute, which is operated by AURA, Inc., under NASA contract NAS5-26555. These observations are associated with program \# GO-13659. 
This work is based on observations made with the NASA/ESA Hubble Space Telescope.
The research made extensive use of NASA's Astrophysics Data System bibliographic services. This research made use of Astropy, a community-developed core Python package for astronomy \citep{Astropy:2013}, NumPy \citep{numpy}, and Matplotlib \citep{Hunter:2007matplotlib}. Additionally, the study made use of two dust extinction tools - \texttt{pyextinction} created by Morgan Fouesneau and hosted at https://github.com/mfouesneau/pyextinction, and the Bayesian Extinction and Stellar Tool created by Karl Gordon \citep[BEAST]{Gordon:2016} and hosted at https://github.com/BEAST-Fitting/beast.

\section{Appendix A: Model Sensitivity Tests}
\label{sec:appendixA}

A number of factors influence the calculated slope of the reddening vector using the red clump method. We perform tests to determine how this slope varies with changing parameters such as:\\\\
$\bullet$ the width of the selection boundary around the red clump, \\
$\bullet$ the intrinsic width of the unreddened red clump, and \\
$\bullet$ the unreddened red clump location.\\

We specify the width of the reddened red clump selection region by first calculating the feature's approximate slope and then adding and subtracting a percent to this slope to define the slopes of the two boundaries tangent and extending redward of the unreddened RC ellipse (see Sec. \ref{sec:SlopeMeasuring}, top). To understand the behavior of the resulting reddening vector slope as we define these tangent boundaries, we vary the percent added and subtracted to the approximate slope between 0\% and 40\% (Fig. \ref{fig:approxSlopeWidthSens}, top). A 0\% variation would mean that the reddened red clump boundary encompasses a relatively narrow region defined by two tangents to the ellipse of the unreddened red clump running parallel to the approximate slope. A 40\% variation, on the other hand, means that the red clump boundary encompasses a relatively large cone-shaped region whose width is set by ellipse tangents with slopes $\pm$ 40\% of the approximate slope. These selection boundaries are illustrated in Fig. \ref{fig:vectorSlopeSens} where we more realistically depict the $\pm$ 5\% case rather than the 0\% case.

We also explore the effect the intrinsic width of the unreddened red clump has on the slope of the reddening vector (Fig. \ref{fig:approxSlopeWidthSens}, middle). This width is present due to the age and metallicity uncertainties of the red clump discussed in Sec \ref{sec:UnredRClocation}. We observe the effect of a changing width on our results by varying $\sigma_{col}$ from 0.06 mags (see Table \ref{tab:unredRCloc}) to twice this value. 

Lastly, to allow for a potential ambiguity in the unreddened position of the red clump in magnitude due to the distance effect discussed in Sec. \ref{sec:UnredRClocation}, we let the centroid of the RC distance modulus vary by 0.15 mags to simulate the effect of a distance shift away from 62 kpc ($\mu=18.96$, \citep{Scowcroft:2016hm}) by \about 4.5 kpc (Fig. \ref{fig:approxSlopeWidthSens}, bottom).

We observe that the variation in both the approximate reddening vector slope and the width of the unreddened red clump produces relatively constant results which fall within the overall uncertainties.  The slope of \textit{F225W} is most susceptible to variation due to the steepness and faintness of the red clump at this wavelength.  Our third sensitivity test which varies the magnitude position of the unreddened red clump by $\pm$ 0.15 mags away from the values given in Table \ref{tab:unredRCloc} results in a calculated reddening vector slope with a similar offset across each CMD.  For the range of magnitude variations we find a $\chi^2_{min}$ which supports a significant line-of-sight depth (between 7 and 15 kpc). We therefore conclude that a variation in the unreddened red clump zero-point does not remove the necessity to account for a substantial galactic depth when analyzing the extinction curve using the red clump method.

\bibliographystyle{yahapj}
\bibliography{dustbib}

\begin{thebibliography}{}
\providecommand\natexlab[1]{#1}
\providecommand\JournalTitle[1]{#1}

\bibitem[{{Alves}(2000)}]{Alves:2000}
{Alves}, D.~R. 2000,
  \href{http://dx.doi.org/10.1086/309278}{\JournalTitle{\apj}, 539, 732}

\bibitem[{{Astropy Collaboration} {et~al.}(2013){Astropy Collaboration},
  {Robitaille}, {Tollerud}, {Greenfield}, {Droettboom}, {Bray}, {Aldcroft},
  {Davis}, {Ginsburg}, {Price-Whelan}, {Kerzendorf}, {Conley}, {Crighton},
  {Barbary}, {Muna}, {Ferguson}, {Grollier}, {Parikh}, {Nair}, {Unther},
  {Deil}, {Woillez}, {Conseil}, {Kramer}, {Turner}, {Singer}, {Fox}, {Weaver},
  {Zabalza}, {Edwards}, {Azalee Bostroem}, {Burke}, {Casey}, {Crawford},
  {Dencheva}, {Ely}, {Jenness}, {Labrie}, {Lim}, {Pierfederici}, {Pontzen},
  {Ptak}, {Refsdal}, {Servillat}, \& {Streicher}}]{Astropy:2013}
{Astropy Collaboration}, {Robitaille}, T.~P., {Tollerud}, E.~J., {et~al.} 2013,
  \href{http://dx.doi.org/10.1051/0004-6361/201322068}{\JournalTitle{\aap},
  558, A33}

\bibitem[{{Bovy} {et~al.}(2014){Bovy}, {Nidever}, {Rix}, {Girardi}, {Zasowski},
  {Chojnowski}, {Holtzman}, {Epstein}, {Frinchaboy}, {Hayden}, {Rodrigues},
  {Majewski}, {Johnson}, {Pinsonneault}, {Stello}, {Allende Prieto}, {Andrews},
  {Basu}, {Beers}, {Bizyaev}, {Burton}, {Chaplin}, {Cunha}, {Elsworth},
  {Garc{\'{\i}}a}, {Garc{\'{\i}}a-Her{\'n}andez}, {Garc{\'{\i}}a P{\'e}rez},
  {Hearty}, {Hekker}, {Kallinger}, {Kinemuchi}, {Koesterke},
  {M{\'e}sz{\'a}ros}, {Mosser}, {O'Connell}, {Oravetz}, {Pan}, {Robin},
  {Schiavon}, {Schneider}, {Schultheis}, {Serenelli}, {Shetrone}, {Silva
  Aguirre}, {Simmons}, {Skrutskie}, {Smith}, {Stassun}, {Weinberg}, {Wilson},
  \& {Zamora}}]{Bovy:2014}
{Bovy}, J., {Nidever}, D.~L., {Rix}, H.-W., {et~al.} 2014,
  \href{http://dx.doi.org/10.1088/0004-637X/790/2/127}{\JournalTitle{\apj},
  790, 127}

\bibitem[{{Bressan} {et~al.}(2012){Bressan}, {Marigo}, {Girardi}, {Salasnich},
  {Dal Cero}, {Rubele}, \& {Nanni}}]{Bressan:2012}
{Bressan}, A., {Marigo}, P., {Girardi}, L., {et~al.} 2012,
  \href{http://dx.doi.org/10.1111/j.1365-2966.2012.21948.x}{\JournalTitle{\mnras},
  427, 127}

\bibitem[{{Caldwell} \& {Coulson}(1986)}]{CaldwellCoulson:1986}
{Caldwell}, J.~A.~R., \& {Coulson}, I.~M. 1986,
  \href{http://dx.doi.org/10.1093/mnras/218.2.223}{\JournalTitle{\mnras}, 218,
  223}

\bibitem[{{Calzetti} {et~al.}(1994){Calzetti}, {Kinney}, \&
  {Storchi-Bergmann}}]{Calzetti:1994}
{Calzetti}, D., {Kinney}, A.~L., \& {Storchi-Bergmann}, T. 1994,
  \href{http://dx.doi.org/10.1086/174346}{\JournalTitle{\apj}, 429, 582}

\bibitem[{{Cannon}(1970)}]{Cannon:1970}
{Cannon}, R.~D. 1970,
  \href{http://dx.doi.org/10.1093/mnras/150.1.111}{\JournalTitle{\mnras}, 150,
  111}

\bibitem[{{Cardelli} {et~al.}(1989){Cardelli}, {Clayton}, \&
  {Mathis}}]{CCM:1989}
{Cardelli}, J.~A., {Clayton}, G.~C., \& {Mathis}, J.~S. 1989,
  \href{http://dx.doi.org/10.1086/167900}{\JournalTitle{\apj}, 345, 245}

\bibitem[{Cardelli {et~al.}(1992)Cardelli, Sembach, \& Mathis}]{Cardelli:1992}
Cardelli, J.~A., Sembach, K.~R., \& Mathis, J.~S. 1992,
  \JournalTitle{Astronomical Journal (ISSN 0004-6256)}, 104, 1916

\bibitem[{{Cartledge} {et~al.}(2005){Cartledge}, {Clayton}, {Gordon},
  {Rachford}, {Draine}, {Martin}, {Mathis}, {Misselt}, {Sofia}, {Whittet}, \&
  {Wolff}}]{Cartledge:2005}
{Cartledge}, S.~I.~B., {Clayton}, G.~C., {Gordon}, K.~D., {et~al.} 2005,
  \href{http://dx.doi.org/10.1086/431922}{\JournalTitle{\apj}, 630, 355}

\bibitem[{{Chen} {et~al.}(2015){Chen}, {Bressan}, {Girardi}, {Marigo}, {Kong},
  \& {Lanza}}]{Chen:2015}
{Chen}, Y., {Bressan}, A., {Girardi}, L., {et~al.} 2015,
  \href{http://dx.doi.org/10.1093/mnras/stv1281}{\JournalTitle{\mnras}, 452,
  1068}

\bibitem[{{Cignoni} {et~al.}(2009){Cignoni}, {Sabbi}, {Nota}, {Tosi},
  {Degl'Innocenti}, {Moroni}, {Angeretti}, {Carlson}, {Gallagher}, {Meixner},
  {Sirianni}, \& {Smith}}]{Cignoni:2009}
{Cignoni}, M., {Sabbi}, E., {Nota}, A., {et~al.} 2009,
  \href{http://dx.doi.org/10.1088/0004-6256/137/3/3668}{\JournalTitle{\aj},
  137, 3668}

\bibitem[{{Clayton} \& {Martin}(1985)}]{ClaytonMartin:1985}
{Clayton}, G.~C., \& {Martin}, P.~G. 1985,
  \href{http://dx.doi.org/10.1086/162821}{\JournalTitle{\apj}, 288, 558}

\bibitem[{{Dalcanton} {et~al.}(2015){Dalcanton}, {Fouesneau}, {Hogg}, {Lang},
  {Leroy}, {Gordon}, {Sandstrom}, {Weisz}, {Williams}, {Bell}, {Dong},
  {Gilbert}, {Gouliermis}, {Guhathakurta}, {Lauer}, {Schruba}, {Seth}, \&
  {Skillman}}]{Dalcanton:2015bl}
{Dalcanton}, J.~J., {Fouesneau}, M., {Hogg}, D.~W., {et~al.} 2015,
  \href{http://dx.doi.org/10.1088/0004-637X/814/1/3}{\JournalTitle{\apj}, 814,
  3}

\bibitem[{{De Marchi} \& {Panagia}(2014)}]{DeMarchi:2014en}
{De Marchi}, G., \& {Panagia}, N. 2014,
  \href{http://dx.doi.org/10.1093/mnras/stu1694}{\JournalTitle{\mnras}, 445,
  93}

\bibitem[{{De Marchi} {et~al.}(2014){De Marchi}, {Panagia}, \&
  {Girardi}}]{DeMarchi:2014fe}
{De Marchi}, G., {Panagia}, N., \& {Girardi}, L. 2014,
  \href{http://dx.doi.org/10.1093/mnras/stt2233}{\JournalTitle{\mnras}, 438,
  513}

\bibitem[{{De Marchi} {et~al.}(2016){De Marchi}, {Panagia}, {Sabbi}, {Lennon},
  {Anderson}, {van der Marel}, {Cignoni}, {Grebel}, {Larsen}, {Zaritsky},
  {Zeidler}, {Gouliermis}, \& {Aloisi}}]{DeMarchi:2016}
{De Marchi}, G., {Panagia}, N., {Sabbi}, E., {et~al.} 2016,
  \href{http://dx.doi.org/10.1093/mnras/stv2528}{\JournalTitle{\mnras}, 455,
  4373}

\bibitem[{{Dolphin}(2000)}]{Dolphin:2000}
{Dolphin}, A.~E. 2000,
  \href{http://dx.doi.org/10.1086/316630}{\JournalTitle{\pasp}, 112, 1383}

\bibitem[{{Dolphin}(2002)}]{Dolphin:2002}
---. 2002,
  \href{http://dx.doi.org/10.1046/j.1365-8711.2002.05271.x}{\JournalTitle{\mnras},
  332, 91}

\bibitem[{Dressel(2014)}]{Dressel:2014}
Dressel, L. 2014,
  \href{http://documents.stsci.edu/hst/acs/documents/handbooks/cycle22/cover.html}{\JournalTitle{"Wide
  Field Camera 3 Instrument Handbook, Version 6.0", (Baltimore: STScI)}}

\bibitem[{{Dufour}(1984)}]{Dufour:1984}
{Dufour}, R.~J. 1984, 108, 353

\bibitem[{{Dunkin} \& {Crawford}(1998)}]{Dunkin:1998}
{Dunkin}, S.~K., \& {Crawford}, I.~A. 1998,
  \href{http://dx.doi.org/10.1046/j.1365-8711.1998.01612.x}{\JournalTitle{\mnras},
  298, 275}

\bibitem[{{Fitzpatrick}(1985)}]{Fitzpatrick:1985}
{Fitzpatrick}, E.~L. 1985,
  \href{http://dx.doi.org/10.1086/163694}{\JournalTitle{\apj}, 299, 219}

\bibitem[{{Fitzpatrick}(1999)}]{Fitzpatrick:1999}
---. 1999, \href{http://dx.doi.org/10.1086/316293}{\JournalTitle{\pasp}, 111,
  63}

\bibitem[{{Florsch} {et~al.}(1981){Florsch}, {Marcout}, \&
  {Fleck}}]{Florsch:1981}
{Florsch}, A., {Marcout}, J., \& {Fleck}, E. 1981, \JournalTitle{\aap}, 96, 158

\bibitem[{{Francis} \& {Anderson}(2014)}]{Francis:2014}
{Francis}, C., \& {Anderson}, E. 2014,
  \href{http://dx.doi.org/10.1093/mnras/stu631}{\JournalTitle{\mnras}, 441,
  1105}

\bibitem[{{Gall} {et~al.}(2014){Gall}, {Hjorth}, {Watson}, {Dwek}, {Maund},
  {Fox}, {Leloudas}, {Malesani}, \& {Day-Jones}}]{Gall:2014}
{Gall}, C., {Hjorth}, J., {Watson}, D., {et~al.} 2014,
  \href{http://dx.doi.org/10.1038/nature13558}{\JournalTitle{\nat}, 511, 326}

\bibitem[{{Galliano} {et~al.}(2005){Galliano}, {Madden}, {Jones}, {Wilson}, \&
  {Bernard}}]{Galliano:2005}
{Galliano}, F., {Madden}, S.~C., {Jones}, A.~P., {Wilson}, C.~D., \& {Bernard},
  J.-P. 2005,
  \href{http://dx.doi.org/10.1051/0004-6361:20042369}{\JournalTitle{\aap}, 434,
  867}

\bibitem[{{Gardiner} \& {Hatzidimitriou}(1992)}]{GardinerHatzidimitriou:1992}
{Gardiner}, L.~T., \& {Hatzidimitriou}, D. 1992,
  \href{http://dx.doi.org/10.1093/mnras/257.2.195}{\JournalTitle{\mnras}, 257,
  195}

\bibitem[{{Gardiner} \& {Hawkins}(1991)}]{Gardiner:1991}
{Gardiner}, L.~T., \& {Hawkins}, M.~R.~S. 1991,
  \href{http://dx.doi.org/10.1093/mnras/251.1.174}{\JournalTitle{\mnras}, 251,
  174}

\bibitem[{{Girardi}(2016)}]{Girardi:2016}
{Girardi}, L. 2016,
  \href{http://dx.doi.org/10.1146/annurev-astro-081915-023354}{\JournalTitle{\araa},
  54, 95}

\bibitem[{{Girardi} \& {Salaris}(2001)}]{GirardiSalaris:2001}
{Girardi}, L., \& {Salaris}, M. 2001,
  \href{http://dx.doi.org/10.1046/j.1365-8711.2001.04084.x}{\JournalTitle{\mnras},
  323, 109}

\bibitem[{{Glatt} {et~al.}(2008){Glatt}, {Gallagher}, {Grebel}, {Nota},
  {Sabbi}, {Sirianni}, {Clementini}, {Tosi}, {Harbeck}, {Koch}, \&
  {Cracraft}}]{Glatt:2008}
{Glatt}, K., {Gallagher}, III, J.~S., {Grebel}, E.~K., {et~al.} 2008,
  \href{http://dx.doi.org/10.1088/0004-6256/135/4/1106}{\JournalTitle{\aj},
  135, 1106}

\bibitem[{{Gordon} \& {Clayton}(1998)}]{Gordon:1998}
{Gordon}, K.~D., \& {Clayton}, G.~C. 1998,
  \href{http://dx.doi.org/10.1086/305774}{\JournalTitle{\apj}, 500, 816}

\bibitem[{{Gordon} {et~al.}(2003){Gordon}, {Clayton}, {Misselt}, {Landolt}, \&
  {Wolff}}]{Gordon:2003}
{Gordon}, K.~D., {Clayton}, G.~C., {Misselt}, K.~A., {Landolt}, A.~U., \&
  {Wolff}, M.~J. 2003,
  \href{http://dx.doi.org/10.1086/376774}{\JournalTitle{\apj}, 594, 279}

\bibitem[{{Gordon} {et~al.}(2016){Gordon}, {Fouesneau}, {Arab}, {Tchernyshyov},
  {Weisz}, {Dalcanton}, {Williams}, {Bell}, {Bianchi}, {Boyer}, {Choi},
  {Dolphin}, {Girardi}, {Hogg}, {Kalirai}, {Kapala}, {Lewis}, {Rix},
  {Sandstrom}, \& {Skillman}}]{Gordon:2016}
{Gordon}, K.~D., {Fouesneau}, M., {Arab}, H., {et~al.} 2016,
  \href{http://dx.doi.org/10.3847/0004-637X/826/2/104}{\JournalTitle{\apj},
  826, 104}

\bibitem[{{Hagen} {et~al.}(2017){Hagen}, {Siegel}, {Hoversten}, {Gronwall},
  {Immler}, \& {Hagen}}]{Hagen:2016smc}
{Hagen}, L.~M.~Z., {Siegel}, M.~H., {Hoversten}, E.~A., {et~al.} 2017,
  \href{http://dx.doi.org/10.1093/mnras/stw2954}{\JournalTitle{\mnras}, 466,
  4540}

\bibitem[{{Haschke} {et~al.}(2012){Haschke}, {Grebel}, \&
  {Duffau}}]{Haschke:2012smc}
{Haschke}, R., {Grebel}, E.~K., \& {Duffau}, S. 2012,
  \href{http://dx.doi.org/10.1088/0004-6256/144/4/107}{\JournalTitle{\aj}, 144,
  107}

\bibitem[{{Hatzidimitriou} \& {Hawkins}(1989)}]{HatzidimHawkins:1989}
{Hatzidimitriou}, D., \& {Hawkins}, M.~R.~S. 1989,
  \href{http://dx.doi.org/10.1093/mnras/241.4.667}{\JournalTitle{\mnras}, 241,
  667}

\bibitem[{{Hennebelle} \& {Falgarone}(2012)}]{Hennebelle:2012}
{Hennebelle}, P., \& {Falgarone}, E. 2012,
  \href{http://dx.doi.org/10.1007/s00159-012-0055-y}{\JournalTitle{\aapr}, 20,
  55}

\bibitem[{{Hill} {et~al.}(2008){Hill}, {Benjamin}, {Kowal}, {Reynolds},
  {Haffner}, \& {Lazarian}}]{Hill:2008}
{Hill}, A.~S., {Benjamin}, R.~A., {Kowal}, G., {et~al.} 2008,
  \href{http://dx.doi.org/10.1086/590543}{\JournalTitle{\apj}, 686, 363}

\bibitem[{Hunter(2007)}]{Hunter:2007matplotlib}
Hunter, J.~D. 2007,
  \href{http://dx.doi.org/10.1109/MCSE.2007.55}{\JournalTitle{Computing in
  Science and Engineering}, 9, 90}

\bibitem[{{Isobe} {et~al.}(1990){Isobe}, {Feigelson}, {Akritas}, \&
  {Babu}}]{Isobe:1990}
{Isobe}, T., {Feigelson}, E.~D., {Akritas}, M.~G., \& {Babu}, G.~J. 1990,
  \href{http://dx.doi.org/10.1086/169390}{\JournalTitle{\apj}, 364, 104}

\bibitem[{{Jacyszyn-Dobrzeniecka} {et~al.}(2016){Jacyszyn-Dobrzeniecka},
  {Skowron}, {Mr{\'o}z}, {Skowron}, {Soszy{\'n}ski}, {Udalski}, {Pietrukowicz},
  {Koz{\l}owski}, {Wyrzykowski}, {Poleski}, {Pawlak}, {Szyma{\'n}ski}, \&
  {Ulaczyk}}]{JD:2016a}
{Jacyszyn-Dobrzeniecka}, A.~M., {Skowron}, D.~M., {Mr{\'o}z}, P., {et~al.}
  2016, \JournalTitle{\actaa}, 66, 149

\bibitem[{{Jacyszyn-Dobrzeniecka} {et~al.}(2017){Jacyszyn-Dobrzeniecka},
  {Skowron}, {Mr{\'o}z}, {Soszy{\'n}ski}, {Udalski}, {Pietrukowicz}, {Skowron},
  {Poleski}, {Koz{\l}owski}, {Wyrzykowski}, {Pawlak}, {Szyma{\'n}ski}, \&
  {Ulaczyk}}]{JD:2016b}
---. 2017, \JournalTitle{\actaa}, 67, 1

\bibitem[{{Kainulainen} {et~al.}(2009){Kainulainen}, {Beuther}, {Henning}, \&
  {Plume}}]{Kainulainen:2009}
{Kainulainen}, J., {Beuther}, H., {Henning}, T., \& {Plume}, R. 2009,
  \href{http://dx.doi.org/10.1051/0004-6361/200913605}{\JournalTitle{\aap},
  508, L35}

\bibitem[{{Kapakos} \& {Hatzidimitriou}(2012)}]{Kapakos:2012}
{Kapakos}, E., \& {Hatzidimitriou}, D. 2012,
  \href{http://dx.doi.org/10.1111/j.1365-2966.2012.21834.x}{\JournalTitle{\mnras},
  426, 2063}

\bibitem[{{Kroupa}(2001)}]{Kroupa:2001}
{Kroupa}, P. 2001,
  \href{http://dx.doi.org/10.1046/j.1365-8711.2001.04022.x}{\JournalTitle{\mnras},
  322, 231}

\bibitem[{{Kurt} {et~al.}(1999){Kurt}, {Dufour}, {Garnett}, {Skillman},
  {Mathis}, {Peimbert}, {Torres-Peimbert}, \& {Ruiz}}]{Kurt:1999}
{Kurt}, C.~M., {Dufour}, R.~J., {Garnett}, D.~R., {et~al.} 1999,
  \href{http://dx.doi.org/10.1086/307271}{\JournalTitle{\apj}, 518, 246}

\bibitem[{{Lee} {et~al.}(2005){Lee}, {Rolleston}, {Dufton}, \&
  {Ryans}}]{Lee:2005}
{Lee}, J.-K., {Rolleston}, W.~R.~J., {Dufton}, P.~L., \& {Ryans}, R.~S.~I.
  2005,
  \href{http://dx.doi.org/10.1051/0004-6361:20041345}{\JournalTitle{\aap}, 429,
  1025}

\bibitem[{{Lequeux} {et~al.}(1982){Lequeux}, {Maurice}, {Prevot-Burnichon},
  {Prevot}, \& {Rocca-Volmerange}}]{Lequeux:1982}
{Lequeux}, J., {Maurice}, E., {Prevot-Burnichon}, M.-L., {Prevot}, L., \&
  {Rocca-Volmerange}, B. 1982, \JournalTitle{\aap}, 113, L15

\bibitem[{{Ma{\'{\i}}z Apell{\'a}niz} \& {Rubio}(2012)}]{MaizApellaniz:2012}
{Ma{\'{\i}}z Apell{\'a}niz}, J., \& {Rubio}, M. 2012,
  \href{http://dx.doi.org/10.1051/0004-6361/201118712}{\JournalTitle{\aap},
  541, A54}

\bibitem[{{Ma{\'{\i}}z Apell{\'a}niz} {et~al.}(2014){Ma{\'{\i}}z
  Apell{\'a}niz}, {Evans}, {Barb{\'a}}, {Gr{\"a}fener}, {Bestenlehner},
  {Crowther}, {Garc{\'{\i}}a}, {Herrero}, {Sana}, {Sim{\'o}n-D{\'{\i}}az},
  {Taylor}, {van Loon}, {Vink}, \& {Walborn}}]{MA:2014}
{Ma{\'{\i}}z Apell{\'a}niz}, J., {Evans}, C.~J., {Barb{\'a}}, R.~H., {et~al.}
  2014,
  \href{http://dx.doi.org/10.1051/0004-6361/201423439}{\JournalTitle{\aap},
  564, A63}

\bibitem[{{Massa} {et~al.}(1983){Massa}, {Savage}, \&
  {Fitzpatrick}}]{Massa:1983}
{Massa}, D., {Savage}, B.~D., \& {Fitzpatrick}, E.~L. 1983,
  \href{http://dx.doi.org/10.1086/160813}{\JournalTitle{\apj}, 266, 662}

\bibitem[{{Monson} {et~al.}(2012){Monson}, {Freedman}, {Madore}, {Persson},
  {Scowcroft}, {Seibert}, \& {Rigby}}]{Monson:2012}
{Monson}, A.~J., {Freedman}, W.~L., {Madore}, B.~F., {et~al.} 2012,
  \href{http://dx.doi.org/10.1088/0004-637X/759/2/146}{\JournalTitle{\apj},
  759, 146}

\bibitem[{{Muller} {et~al.}(2003){Muller}, {Staveley-Smith}, {Zealey}, \&
  {Stanimirovi{\'c}}}]{Muller:2003}
{Muller}, E., {Staveley-Smith}, L., {Zealey}, W., \& {Stanimirovi{\'c}}, S.
  2003,
  \href{http://dx.doi.org/10.1046/j.1365-8711.2003.06147.x}{\JournalTitle{\mnras},
  339, 105}

\bibitem[{{Nataf} {et~al.}(2013){Nataf}, {Gould}, {Fouqu{\'e}}, {Gonzalez},
  {Johnson}, {Skowron}, {Udalski}, {Szyma{\'n}ski}, {Kubiak},
  {Pietrzy{\'n}ski}, {Soszy{\'n}ski}, {Ulaczyk}, {Wyrzykowski}, \&
  {Poleski}}]{Nataf:2013}
{Nataf}, D.~M., {Gould}, A., {Fouqu{\'e}}, P., {et~al.} 2013,
  \href{http://dx.doi.org/10.1088/0004-637X/769/2/88}{\JournalTitle{\apj}, 769,
  88}

\bibitem[{{Nidever} {et~al.}(2013){Nidever}, {Monachesi}, {Bell}, {Majewski},
  {Mu{\~n}oz}, \& {Beaton}}]{Nidever:2013ji}
{Nidever}, D.~L., {Monachesi}, A., {Bell}, E.~F., {et~al.} 2013,
  \href{http://dx.doi.org/10.1088/0004-637X/779/2/145}{\JournalTitle{\apj},
  779, 145}

\bibitem[{{Noll} \& {Pierini}(2005)}]{Noll:2005}
{Noll}, S., \& {Pierini}, D. 2005,
  \href{http://dx.doi.org/10.1051/0004-6361:20053635}{\JournalTitle{\aap}, 444,
  137}

\bibitem[{{Paczy{\'n}ski} \& {Stanek}(1998)}]{Paczynski:1998}
{Paczy{\'n}ski}, B., \& {Stanek}, K.~Z. 1998,
  \href{http://dx.doi.org/10.1086/311181}{\JournalTitle{\apjl}, 494, L219}

\bibitem[{{Percival} \& {Salaris}(2003)}]{Percival:2003}
{Percival}, S.~M., \& {Salaris}, M. 2003,
  \href{http://dx.doi.org/10.1046/j.1365-8711.2003.06691.x}{\JournalTitle{\mnras},
  343, 539}

\bibitem[{{Prevot} {et~al.}(1984){Prevot}, {Lequeux}, {Prevot}, {Maurice}, \&
  {Rocca-Volmerange}}]{Prevot:1984}
{Prevot}, M.~L., {Lequeux}, J., {Prevot}, L., {Maurice}, E., \&
  {Rocca-Volmerange}, B. 1984, \JournalTitle{\aap}, 132, 389

\bibitem[{{Rolleston} {et~al.}(1999){Rolleston}, {Dufton}, {McErlean}, \&
  {Venn}}]{Rolleston:1999}
{Rolleston}, W.~R.~J., {Dufton}, P.~L., {McErlean}, N.~D., \& {Venn}, K.~A.
  1999, \JournalTitle{\aap}, 348, 728

\bibitem[{{Rolleston} {et~al.}(2003){Rolleston}, {Venn}, {Tolstoy}, \&
  {Dufton}}]{Rolleston:2003}
{Rolleston}, W.~R.~J., {Venn}, K., {Tolstoy}, E., \& {Dufton}, P.~L. 2003,
  \href{http://dx.doi.org/10.1051/0004-6361:20021653}{\JournalTitle{\aap}, 400,
  21}

\bibitem[{{Rubele} {et~al.}(2015){Rubele}, {Girardi}, {Kerber}, {Cioni},
  {Piatti}, {Zaggia}, {Bekki}, {Bressan}, {Clementini}, {de Grijs}, {Emerson},
  {Groenewegen}, {Ivanov}, {Marconi}, {Marigo}, {Moretti}, {Ripepi},
  {Subramanian}, {Tatton}, \& {van Loon}}]{Rubele:2015vmc}
{Rubele}, S., {Girardi}, L., {Kerber}, L., {et~al.} 2015,
  \href{http://dx.doi.org/10.1093/mnras/stv141}{\JournalTitle{\mnras}, 449,
  639}

\bibitem[{{Russell} \& {Dopita}(1992)}]{RussellDopita:1992}
{Russell}, S.~C., \& {Dopita}, M.~A. 1992,
  \href{http://dx.doi.org/10.1086/170893}{\JournalTitle{\apj}, 384, 508}

\bibitem[{{Sabbi} {et~al.}(2009){Sabbi}, {Gallagher}, {Tosi}, {Anderson},
  {Nota}, {Grebel}, {Cignoni}, {Cole}, {Da Costa}, {Harbeck}, {Glatt}, \&
  {Marconi}}]{Sabbi:2009}
{Sabbi}, E., {Gallagher}, J.~S., {Tosi}, M., {et~al.} 2009,
  \href{http://dx.doi.org/10.1088/0004-637X/703/1/721}{\JournalTitle{\apj},
  703, 721}

\bibitem[{{Sabbi} {et~al.}(2013){Sabbi}, {Anderson}, {Lennon}, {van der Marel},
  {Aloisi}, {Boyer}, {Cignoni}, {de Marchi}, {de Mink}, {Evans}, {Gallagher},
  {Gordon}, {Gouliermis}, {Grebel}, {Koekemoer}, {Larsen}, {Panagia}, {Ryon},
  {Smith}, {Tosi}, \& {Zaritsky}}]{Sabbi:2013}
{Sabbi}, E., {Anderson}, J., {Lennon}, D.~J., {et~al.} 2013,
  \href{http://dx.doi.org/10.1088/0004-6256/146/3/53}{\JournalTitle{\aj}, 146,
  53}

\bibitem[{{Sabbi} {et~al.}(2016){Sabbi}, {Lennon}, {Anderson}, {Cignoni}, {van
  der Marel}, {Zaritsky}, {De Marchi}, {Panagia}, {Gouliermis}, {Grebel},
  {Gallagher}, {Smith}, {Sana}, {Aloisi}, {Tosi}, {Evans}, {Arab}, {Boyer}, {de
  Mink}, {Gordon}, {Koekemoer}, {Larsen}, {Ryon}, \& {Zeidler}}]{Sabbi:2016}
{Sabbi}, E., {Lennon}, D.~J., {Anderson}, J., {et~al.} 2016,
  \href{http://dx.doi.org/10.3847/0067-0049/222/1/11}{\JournalTitle{\apjs},
  222, 11}

\bibitem[{{Schlafly} {et~al.}(2016){Schlafly}, {Meisner}, {Stutz},
  {Kainulainen}, {Peek}, {Tchernyshyov}, {Rix}, {Finkbeiner}, {Covey}, {Green},
  {Bell}, {Burgett}, {Chambers}, {Draper}, {Flewelling}, {Hodapp}, {Kaiser},
  {Magnier}, {Martin}, {Metcalfe}, {Wainscoat}, \& {Waters}}]{Schlafly:2016}
{Schlafly}, E.~F., {Meisner}, A.~M., {Stutz}, A.~M., {et~al.} 2016,
  \href{http://dx.doi.org/10.3847/0004-637X/821/2/78}{\JournalTitle{\apj}, 821,
  78}

\bibitem[{{Scowcroft} {et~al.}(2016){Scowcroft}, {Freedman}, {Madore},
  {Monson}, {Persson}, {Rich}, {Seibert}, \& {Rigby}}]{Scowcroft:2016hm}
{Scowcroft}, V., {Freedman}, W.~L., {Madore}, B.~F., {et~al.} 2016,
  \href{http://dx.doi.org/10.3847/0004-637X/816/2/49}{\JournalTitle{\apj}, 816,
  49}

\bibitem[{{Strom} {et~al.}(1971){Strom}, {Strom}, \& {Yost}}]{Strom:1971}
{Strom}, K.~M., {Strom}, S.~E., \& {Yost}, J. 1971,
  \href{http://dx.doi.org/10.1086/150915}{\JournalTitle{\apj}, 165, 479}

\bibitem[{{Subramanian} \& {Subramaniam}(2009)}]{Subramanian:2009}
{Subramanian}, S., \& {Subramaniam}, A. 2009,
  \href{http://dx.doi.org/10.1051/0004-6361/200811029}{\JournalTitle{\aap},
  496, 399}

\bibitem[{{Subramanian} \& {Subramaniam}(2012)}]{Subramanian:2012}
---. 2012,
  \href{http://dx.doi.org/10.1088/0004-637X/744/2/128}{\JournalTitle{\apj},
  744, 128}

\bibitem[{{Subramanian} {et~al.}(2017){Subramanian}, {Rubele}, {Sun},
  {Girardi}, {de Grijs}, {van Loon}, {Cioni}, {Piatti}, {Bekki}, {Emerson},
  {Ivanov}, {Kerber}, {Marconi}, {Ripepi}, \& {Tatton}}]{Subramanian:2017}
{Subramanian}, S., {Rubele}, S., {Sun}, N.-C., {et~al.} 2017,
  \href{http://dx.doi.org/10.1093/mnras/stx205}{\JournalTitle{\mnras}, 467,
  2980}

\bibitem[{{Tang} {et~al.}(2014){Tang}, {Bressan}, {Rosenfield}, {Slemer},
  {Marigo}, {Girardi}, \& {Bianchi}}]{Tang:2014}
{Tang}, J., {Bressan}, A., {Rosenfield}, P., {et~al.} 2014,
  \href{http://dx.doi.org/10.1093/mnras/stu2029}{\JournalTitle{\mnras}, 445,
  4287}

\bibitem[{{Trumpler}(1930)}]{Trumpler:1930}
{Trumpler}, R.~J. 1930,
  \href{http://dx.doi.org/10.1086/124039}{\JournalTitle{\pasp}, 42, 214}

\bibitem[{Van Der~Walt {et~al.}(2011)Van Der~Walt, Colbert, \&
  Varoquaux}]{numpy}
Van Der~Walt, S., Colbert, S.~C., \& Varoquaux, G. 2011,
  \href{http://dx.doi.org/10.1109/MCSE.2011.37}{\JournalTitle{Computing in
  Science and Engineering}, 13, 22}

\bibitem[{{Weingartner} \& {Draine}(2001)}]{WD:2001}
{Weingartner}, J.~C., \& {Draine}, B.~T. 2001,
  \href{http://dx.doi.org/10.1086/318651}{\JournalTitle{\apj}, 548, 296}

\bibitem[{{Weisz} {et~al.}(2013){Weisz}, {Dolphin}, {Skillman}, {Holtzman},
  {Dalcanton}, {Cole}, \& {Neary}}]{Weisz:2013mc}
{Weisz}, D.~R., {Dolphin}, A.~E., {Skillman}, E.~D., {et~al.} 2013,
  \href{http://dx.doi.org/10.1093/mnras/stt165}{\JournalTitle{\mnras}, 431,
  364}

\bibitem[{{Welch} {et~al.}(1987){Welch}, {McLaren}, {Madore}, \&
  {McAlary}}]{Welch:1987}
{Welch}, D.~L., {McLaren}, R.~A., {Madore}, B.~F., \& {McAlary}, C.~W. 1987,
  \href{http://dx.doi.org/10.1086/165622}{\JournalTitle{\apj}, 321, 162}

\bibitem[{{Welty} {et~al.}(2012){Welty}, {Xue}, \& {Wong}}]{Welty:2012}
{Welty}, D.~E., {Xue}, R., \& {Wong}, T. 2012,
  \href{http://dx.doi.org/10.1088/0004-637X/745/2/173}{\JournalTitle{\apj},
  745, 173}

\bibitem[{{Williams} {et~al.}(2014){Williams}, {Lang}, {Dalcanton}, {Dolphin},
  {Weisz}, {Bell}, {Bianchi}, {Byler}, {Gilbert}, {Girardi}, {Gordon},
  {Gregersen}, {Johnson}, {Kalirai}, {Lauer}, {Monachesi}, {Rosenfield},
  {Seth}, \& {Skillman}}]{Williams:2014}
{Williams}, B.~F., {Lang}, D., {Dalcanton}, J.~J., {et~al.} 2014,
  \href{http://dx.doi.org/10.1088/0067-0049/215/1/9}{\JournalTitle{\apjs}, 215,
  9}

\end{thebibliography}
\end{document}